\documentstyle[12pt,aaspp4]{article}
\begin{document}

\slugcomment{Astronomical Journal, in press (Jan 2001)}

\title{High-Redshift Quasars Found in Sloan Digital
Sky Survey Commissioning Data III:
A Color Selected Sample at $i^*<20$ in the Fall Equatorial Stripe$^1$}

\author{Xiaohui Fan\altaffilmark{\ref{Princeton},\ref{IAS}},
Michael A. Strauss\altaffilmark{\ref{Princeton}},
Gordon T. Richards\altaffilmark{\ref{Chicago},\ref{PennState}},
Jeffrey A. Newman\altaffilmark{\ref{UCB}},
Robert H. Becker\altaffilmark{\ref{UCDavis},\ref{IGPP}},
Donald P. Schneider\altaffilmark{\ref{PennState}},
James E. Gunn\altaffilmark{\ref{Princeton}},
Marc Davis\altaffilmark{\ref{UCB}},
Richard L. White\altaffilmark{\ref{STScI}},
Robert H. Lupton\altaffilmark{\ref{Princeton}},
John E. Anderson, Jr.\altaffilmark{\ref{Fermilab}},
James Annis\altaffilmark{\ref{Fermilab}},
Neta A. Bahcall\altaffilmark{\ref{Princeton}},
Robert J. Brunner\altaffilmark{\ref{Caltech}},
Istvan Csabai\altaffilmark{\ref{JHU}},
Mamoru Doi\altaffilmark{\ref{UTokyo}},
Masataka Fukugita\altaffilmark{\ref{CosmicRay},\ref{IAS}},
G. S. Hennessy\altaffilmark{\ref{USNO}},
Robert B. Hindsley\altaffilmark{\ref{NRL}},
\v{Z}eljko Ivezi\'{c}\altaffilmark{\ref{Princeton}},
Gillian R. Knapp\altaffilmark{\ref{Princeton}},
Timothy A. McKay\altaffilmark{\ref{Michigan}},
Jeffrey A. Munn\altaffilmark{\ref{Flagstaff}},
Jeffrey R. Pier\altaffilmark{\ref{Flagstaff}},
Alexander S. Szalay\altaffilmark{\ref{JHU}},
and Donald G. York\altaffilmark{\ref{Chicago}}
}

\altaffiltext{1}{Based on observations obtained with the
Sloan Digital Sky Survey, with the Apache Point Observatory
3.5-meter telescope, which is owned and operated by the Astrophysical
Research Consortium, 
by the W. M. Keck Observatory, which is operated as a scientific
partnership among the California Institute of Technology,
the University of California, and NASA,
and was made possible by the generous financial support of the W. M. Keck
Foundation,
and with the Hobby-Eberly Telescope, 
which is a joint project of the University of Texas at Austin,
the Pennsylvania State University, Stanford University,
Ludwig-Maximillians-Universit\"at M\"unchen, and Georg-August-Universit\"at
G\"ottingen.}
\newcounter{address}
\setcounter{address}{2}
\altaffiltext{\theaddress}{Princeton University Observatory, Princeton,
NJ 08544
\label{Princeton}}
\addtocounter{address}{1}
\altaffiltext{\theaddress}{Institute for Advanced Study, Olden Lane,
Princeton, NJ 08540
\label{IAS}}
\addtocounter{address}{1}
\altaffiltext{\theaddress}{University of Chicago, Astronomy \& Astrophysics Center, 5640 S. Ellis Ave., Chicago, IL 60637
\label{Chicago}}
\addtocounter{address}{1}
\altaffiltext{\theaddress}{Department of Astronomy and Astrophysics,
The Pennsylvania State University,
University Park, PA 16802
\label{PennState}}
\addtocounter{address}{1}
\altaffiltext{\theaddress}{Department of Astronomy, University of California, Berkeley, CA 94720-3411
\label{UCB}}
\addtocounter{address}{1}
\altaffiltext{\theaddress}{Physics Dept., University of California, Davis, CA 95616 
\label{UCDavis}}
\addtocounter{address}{1}
\altaffiltext{\theaddress}{IGPP/Lawrence Livermore National Laboratory 
\label{IGPP}}
\addtocounter{address}{1}
\altaffiltext{\theaddress}{Space Telescope Science Institute, Baltimore, MD 21218
\label{STScI}}
\addtocounter{address}{1}
\altaffiltext{\theaddress}{Fermi National Accelerator Laboratory, P.O. Box 500,
Batavia, IL 60510
\label{Fermilab}}
\addtocounter{address}{1}
\altaffiltext{\theaddress}{
Department of Astronomy, California Institute of Technology,
Pasadena, CA 91125
\label{Caltech}}
\addtocounter{address}{1}
\addtocounter{address}{1}
\altaffiltext{\theaddress}{
Department of Physics and Astronomy, The Johns Hopkins University,
   3701 San Martin Drive, Baltimore, MD 21218, USA
\label{JHU}}
\addtocounter{address}{1}
\altaffiltext{\theaddress}{Department of Astronomy and Research Center for the
 Early Universe,
        School of Science, University of Tokyo, Hongo, Bunkyo,
Tokyo, 113-0033 Japan
\label{UTokyo}}
\addtocounter{address}{1}
\altaffiltext{\theaddress}{Institute for Cosmic Ray Research, University of
Tokyo, Midori, Tanashi, Tokyo 188-8502, Japan
\label{CosmicRay}}
\addtocounter{address}{1}
\altaffiltext{\theaddress}{U.S. Naval Observatory,
3450 Massachusetts Ave., NW,
Washington, DC  20392-5420
\label{USNO}}
\addtocounter{address}{1}
\altaffiltext{\theaddress}{Remote Sensing Division, Code 7215, Naval
  Research Laboratory, 4555 Overlook Ave. SW, Washington, DC 20375
\label{NRL}}
\addtocounter{address}{1}
\altaffiltext{\theaddress}{University of Michigan, Department of Physics,
        500 East University, Ann Arbor, MI 48109
\label{Michigan}}
\addtocounter{address}{1}
\altaffiltext{\theaddress}{U.S. Naval Observatory, Flagstaff Station,
P.O. Box 1149,
Flagstaff, AZ  86002-1149
\label{Flagstaff}}

\begin{abstract}
 This is the third paper in a series aimed at
finding high-redshift quasars from five-color ($u'g'r'i'z'$)
imaging data taken along the Celestial Equator by
the Sloan Digital Sky Survey (SDSS) during its commissioning phase.
In this paper, we first present the observations of 
14 bright high-redshift quasars ($3.66 \leq z \leq 4.77$,
$ i^* \lesssim 20$) discovered in the SDSS Fall Equatorial Stripe, 
and the SDSS photometry of two previously known high-redshift quasars in the
same region of the sky.
Combined with the quasars presented in Fan et al. (1999) and Schneider et al. (2000b), 
we define a color-selected flux-limited sample of 39 quasars at $3.6 < z < 5.0$ 
and $i^* \lesssim 20$, covering a total effective area of 182 deg$^2$.
From this sample, we estimate the average spectral power law slope 
in the rest-frame ultraviolet for
quasars at $z\sim4$ to be $-0.79$ with a standard deviation of 0.34,
and the average rest-frame equivalent width of the Ly$\alpha$+N V emission line
to be 69 \AA\ with a standard deviation of 18 \AA.

The selection completeness of this multicolor sample is determined
from the model colors of high-redshift quasars, 
taking into account the distributions of emission line strengths, 
intrinsic continuum slope, the  line and continuum absorption from intervening
material, and the effects of photometric errors.
The average completeness of this sample is about 75\%.
The selection function calculated in this paper will be used to
correct the incompleteness of this color-selected sample and to
derive the high-redshift quasar luminosity function in a subsequent paper.

In the Appendix, we present the observations of an additional 18 
faint quasars ($3.57 \leq z \leq 4.80$, $20.1 < i^* < 20.8$)
discovered in the region on the sky that has been imaged twice.
Several quasars presented in this paper exhibit interesting properties,
including a radio-loud quasar at $z=4.77$, and a narrow-line quasar
(FWHM = 1500 km s$^{-1}$) at $z=3.57$.

\end{abstract}

\section{Introduction}

This paper is the third in a series presenting high-redshift ($z>3.6$)
quasars selected from the commissioning data of the
Sloan Digital Sky Survey (SDSS;
\cite{York00}).
In Papers I and II (Fan et al. 1999a, 2000a) in this series, 
and in Schneider et al. (2000a,~b, 2001), 
we have presented the discovery of $\sim 60$ quasars at $z\gtrsim 3.6$,
selected from $\approx 400 $ deg$^2$ of SDSS multicolor imaging
data along the Celestial Equator.
In this paper, we describe observations
of additional quasar candidates selected in a similar manner
in the Fall Equatorial Stripe, overlapping the areas covered
by \cite{PaperI} and \cite{HET2},
together defining a complete color-selected 
sample of bright high-redshift quasars covering 182 deg$^2$. 
In addition, we search for fainter high-redshift 
quasars based on areas of the sky that have been imaged twice by the SDSS. 

In \S 2, we describe the photometric observations and quasar candidate
selection. 
In \S 3, we present  
the photometry and the spectroscopic observations of 14 newly discovered quasars in the
Fall Equatorial Stripe  at $z\gtrsim 3.6$ and $i^* \lesssim 20$, plus the observations
of two previously known quasars in the same area of the sky.
In this section, we also  develop a method of using the measurements of emission line equivalent widths
and broad-band multicolor photometry to estimate the slope of the intrinsic
power-law continuum.
In \S4, we present the final flux-limited color-selected high-redshift 
quasar sample, which consists of 39 quasars 
at $i^* \lesssim 20$ and $3.6 < z < 5.0$ and covering 182 deg$^2$.
The selection completeness of this sample is calculated in \S5, based on 
a Monte-Carlo simulation of quasar colors as a function of redshift,
using the quasar spectral model of  \cite{FanCss99}.
The effects of redshift, luminosity, continuum shape and
emission line strength on the completeness of the survey are discussed
in detail in \S5.
The results are summarized in \S6.
This color-selected sample is used to study the evolution of the quasar
luminosity function at high redshift in a subsequent paper (Paper IV, 
\cite{Paper4}).

In Appendix A, we provide  updated information on the SDSS filter
response functions, which are used to  calculate quasar model colors and
the selection function.
Part of the Fall Equatorial Stripe area  has been imaged multiple times during the
commissioning phase, allowing a search for quasars fainter than $i^* \sim 20$.
In Appendix B, we present the photometry and spectroscopy of 18 faint
quasars discovered from these data.
They do not yet form a complete sample with well-defined selection criteria.

\section{Photometric Observations and Quasar Selection}

The SDSS telescope (\cite{Siegmund00}),
imaging camera (\cite{Gunnetal}),
and photometric data reduction (\cite{Luptonetal00})
are described in York et al. (2000) (see also Paper I).
Briefly, the telescope, located
at Apache Point Observatory in southeastern New Mexico, has a 2.5m
primary mirror and a wide, essentially distortion-free field.  The
imaging camera contains thirty $2048 \times 2048$ photometric CCDs, which
simultaneously observe 6 parallel $13'$ wide swaths, or {\em scanlines}
of the sky, in 5 broad filters ($u'$, $g'$, $r'$, $i'$, and $z'$) covering
the entire optical band from the atmospheric cutoff in the blue to
the silicon sensitivity cutoff in the red (\cite{F96}, see also Appendix A).
The photometric data are taken in time-delay and integrate (TDI, or
``drift-scan'') mode at sidereal rate;
a given point on the sky passes through each of the five filters in
succession.
The total integration time per filter is 54.1 seconds.
The data were calibrated photometrically by observing secondary standards
in the survey area using a 50cm telescope at Apache Point Observatory. 
The SDSS primary standard star network had not yet been finalized at the time of
these observations,
so the photometric calibration used in this paper is only accurate to 
$\sim$ 10\%, 2\%, 2\%, 2\% and 7\%, in the $u'$, $g'$, $r'$, $i'$, and $z'$ bands,
respectively. 
Thus as in Papers I and II, we will denote the preliminary SDSS magnitudes
presented here as $u^*$, $g^*$, $r^*$, $i^*$ and $z^*$, rather than
the notation $u'$, $g'$, $r'$, $i'$ and $z'$ that will be used for the
final SDSS photometry, which will have photometric accuracy 
of $\lesssim 2\%$ for bright sources.

In this paper, we select bright ($i^* \lesssim 20$) high-redshift
quasar candidates from  
two SDSS imaging runs (Run 94 and 125, see Table B1) in the Fall Equatorial Stripe
at high Galactic latitude ($22^h20^m \lesssim \alpha_{2000} \lesssim 4^h30^m$).
These two runs form a filled {\em stripe} 2.5 degrees wide in declination, 
centered on the Celestial Equator. 
Some quasars found from these two runs have already
been presented in Paper I  and in Schneider et al. (2000b).
The photometric data reduction procedures are described in Paper I. 
In this paper, we use the photometric catalogs produced by 
the most recent version of the SDSS photometric pipeline,
which now corrects for the time 
and spatial variation of the Point Spread Function
(PSF) within a given $10' \times 13'$ CCD frame, 
improving both the photometric accuracy
and the star-galaxy separation.
The 5-$\sigma$ limiting magnitudes are similar to those of Papers I and II,
roughly 22.3, 22.6, 22.7, 22.4 and 20.5 in $u^*$, $g^*$, $r^*$, $i^*$, and
$z^*$, respectively, for a single observation.
The photometry from the new reduction has changed slightly from the 
those in Paper I, but the differences are  almost always smaller
than the quoted photometric errors.
Figure 1 presents the color-color diagrams for all stellar sources
at $i^* < 20.2$ in 25 deg$^2$ of the SDSS imaging data.
The inner parts of the diagrams are shown as contours, linearly spaced in
density of stars per unit area on the color-color diagram.
As in Papers I and II, only sources that are detected in all  three
relevant bands at more than 5 $\sigma$ are plotted. 
The median tracks of quasar colors as a function of redshift from \cite{FanCss99}
are also displayed in Figure 1.

High-redshift quasar candidates were selected using color cuts similar
to those presented in Papers I and II. The final criteria we use to 
define the color-selected complete sample in \S 4 are based on the
colors {\em after} correcting for Galactic
extinction using the reddening map of \cite{Schlegel98}. 
The criteria are as follows: 

1. $gri$ candidates, selected principally from the $g^*-r^*, r^*-i^*$ diagram:

\begin{equation}
 \begin{array}{l}
        (a)\ i^* < 20.05 \\
        (b)\ u^* - g^* > 2.00 \mbox{ or } u^* > 21.00 \\
        (c)\ g^* - r^* > 1.00 \\
        (d)\ r^* - i^* < 0.42 (g^* - r^*) - 0.31 \mbox{ or } g^*- r^* >2.30 \\
        (e)\ i^* - z^* < 0.25
        \end{array}
\end{equation}

2. $riz$ candidates, selected principally from the $r^*-i^*, i^*-z^*$ diagram:

\begin{equation}
\begin{array}{l}
        (a)\ i^* < 20.20  \\
        (b)\ u^* > 22.00 \\
        (c)\ g^* > 22.60 \\
        (d)\ r^* - i^* > 1.00 \\
        (e)\ i^* - z^* < 0.47 (r^* - i^*) - 0.48
        \end{array}
\end{equation}

The intersections of those color cuts with the $g^* - r^*, r^*-i^*$ and
$r^*-i^*,i^*-z^*$ diagrams are illustrated in Figure 1.
The final criteria are more restrictive  than those 
used in Papers I and II, in the sense
that the cuts are further away from the stellar locus. 
Because of the limited observing time for spectroscopic follow-up
observations, we used this tighter cut to limit the number of  
contaminants among the candidates
 in order to complete the sample in the Fall 1999 observing season. 
This modification increases the observing efficiency at the cost of 
lower selection completeness.
The completeness function of the
color selection criteria, and the effects of different
selection constraints on the sample are presented in \S 5. 
In the SDSS main spectroscopic survey, the quasar candidates are selected   
in multi-dimensional color space using a more complicated algorithm
(Newberg et al. 2000,  in preparation, see also \cite{Newberg97}).

 Paper I presented the photometry of two previously known quasars from Run 94 data. 
In Run 125 (in an area not covered by Paper I), two additional quasars are recovered.
SDSSp J005922.65+000301.4\footnote{The naming convention for the SDSS sources is SDSSp JHHMMSS.SS$\pm$DDMMSS.S,
where ``p'' stands for the preliminary SDSS astrometry, and the positions
are expressed in J2000.0 coordinates.
The astrometry is accurate to better than $0.2''$
in each axis.}
is reported in \cite{K95} 
as PSS 0059+0003 ($z=4.16$).
No information on the other quasar, 
SDSSp J024457.19--010809.9, 
has been published in the literature, but it  is reported at the DPOSS website\footnote{\texttt{http://astro.caltech.edu/$^\sim$george/z4.qso}} as
PSS 0244--0108. 
We obtained its spectrum in order to measure its spectral properties. 

\section{Spectroscopic Observations}

 Thirty-three quasar candidates that meet the criteria described
in \S 2 were observed
in the Fall 1999 observing season using the Keck II telescope and the Hobby-Eberly 
Telescope (HET).  Twenty-five of the candidates are identified as 
high-redshift quasars at $z>3.6$. 
The  HET observations of 11 new quasars are presented in a separate
paper (Schneider et al. 2000b). Here we present the Keck observations.

 Spectra of 12 SDSS high-redshift quasars were taken with the Keck II telescope,
using the Low Resolution Imaging Spectrograph (LRIS, \cite{LRIS}) on the nights of
1999 October 15 and 16. A 300 line mm$^{-1}$ grating was used. 
Two quasars (SDSSp J012700.69$-$004559.1 and SDSSp J021043.17$-$001818.4)
were observed with Keck II/LRIS  on 1999 July 21, 
when a 400 line mm$^{-1}$ grating was used.
The exposure times ranged from 600~s to 1200~s.
Observations of the spectrophotometric standard Feige 110 
(\cite{Massey1}, \cite{Massey2}) were used to provide flux
calibrations and to correct for
telluric absorption, and observations of Hg-Ne-Ar lamps were used for wavelength
calibration. The spectra were reduced with standard IRAF procedures.
The calibrated spectra extend from 4000 \AA\ to 9000 \AA\ for
the 300 line mm$^{-1}$ grating spectra, 
with a dispersion of 2.4 \AA/pixel.
Note that  at 8000 \AA$ < \lambda < 9000$ \AA, the spectrum is slightly affected
by second-order contamination from 4000 $ < \lambda < 4500$ \AA. 
However, there is little flux at at $\lambda < 4500$\AA\ for
high-redshift quasars due to the absorption systems, 
and we did not  notice any order-overlap effect in the data. 
The second-order effect, however, does somewhat 
affect the standard star observation, 
the flux calibration, and the removal of the telluric bands.
The spectral coverage is 4000 \AA\ to 8000 \AA\ for
the 400 line mm$^{-1}$ grating spectra,
with a dispersion of 1.9 \AA/pixel.
The resolution of the LRIS spectra is $\sim 7$ \AA.
In addition, spectra of SDSSp J024457.19--010809.9  (PSS 0244--0108) were obtained
with the ARC 3.5m telescope using the Double Imaging Spectrograph (DIS)
on the night of 1999 December 28 for a total exposure time of 2400 s. 
The DIS instrument and spectral data reduction procedures
were described in Paper~I. 

 Table 1 gives the positions, SDSS photometry  and redshifts of the 
14 new SDSS high-redshift quasars, 
and the photometric run from which they were selected, 
as well as the SDSS measurements of the two previously-known quasars.
The photometry is expressed in asinh magnitudes
(\cite{Luptitude}, see also Paper I). 
The photometric errors given are only statistical, and 
do not include calibration errors.
Note that all but one objects presented in this paper are
selected from Run 125. 
Most of the candidates from Run 94 are presented in
Paper I and in \cite{HET2}.
Positions of the 16 confirmed SDSS quasars presented in this paper are
plotted on the color-color diagrams (Figure 1a and 1c) as filled circles. 
Positions of other quasars in the same survey area and non-quasars that
satisfied the selection criteria are also shown in Figure 1.
Finding charts of all objects in Table 1 are given in Figure 2.
They  are $160'' \times 160''$ $i'$ band SDSS images with
an effective exposure time of 54.1 seconds.

We matched the positions of quasars in Table 2 against the FIRST radio survey
(\cite{FIRST}). 
Three of them have FIRST counterparts at 20 cm at the 1 mJy level
(with positional matches better than $1''$):
SDSSp J013108.19+005248.2 ($z=4.19$),
SDSSp J021043.17--001818.4 ($z=4.77$) and SDSSp J030025.23+003224.2 ($z=4.19$),
correspond to unresolved FIRST sources of 
4.00, 9.75 and 7.56 mJy at 20 cm, respectively.
In particular, SDSSp J021043.17--001818.4 ($z=4.77$) is the highest redshift
radio-loud luminous quasar yet known.
None of the objects was detected in the ROSAT full-sky pixel images
(Voges et al. 1999) at a typical 3$\sigma$ upper limit of
$3 \times 10^{-13}$ erg s$^{-1}$ cm$^{-2}$ in the 
0.1 -- 2.4 keV band.

 Figure 3 presents the spectra of the 14 newly discovered SDSS quasars
and of SDSSp J024457.19--010809.9 (PSS 0244--0108).
In the figure, we place the spectra on an absolute flux scale
(to compensate for the uncertainties due to non-photometric conditions
and variable seeing during the night) by forcing the synthetic
$i^*$ magnitudes from the spectra to be the same as
the SDSS photometric measurements.
The emission line properties of the quasars are listed in Table 2.
Central wavelengths and rest frame equivalent widths of five major
emission lines are measured following the procedures of Paper I.

\subsection{Determination of Continuum Properties}
 
 We assume the quasar continuum is a power law with a slope $\alpha$:
$f_{\nu}  \propto \nu^{\alpha}$ (or $f_\lambda \propto \lambda^{-\alpha-2}$).
The continuum magnitude, $AB_{1450}$, 
is defined as the AB magnitude at $\lambda = 1450$ \AA\ in the rest
frame corrected for Galactic extinction. 
AB magnitude can be converted to flux units as:
\begin{equation}
AB_\nu = -2.5 \log f_\nu - 48.60,
\end{equation}
where $f_\nu$ has units of erg s$^{-1}$ cm$^{-2}$ Hz$^{-1}$.
For a power law continuum, $AB_{1450}$ can be converted
to the rest-frame Kron-Cousins B band magnitude:
\begin{equation}
B = AB_{1450} + 2.5 \alpha \log(4400/1450) + 0.12.
\end{equation}
The effective wavelength in the B band is 4400 \AA, and
the factor 0.12 is the zeropoint difference between the AB magnitude system
and the Vega-based system in the B band (\cite{SSG95}).

 Table 3 gives the continuum properties of the quasars. As in Paper I,
redshifts are determined from all emission lines redward of
Ly$\alpha$; Ly$\alpha$ itself is not used, due to absorption from the
Ly$\alpha$ forest on its blue side.
The parameters $\alpha$ and  $AB_{1450}$ can be determined
from the spectrum alone. 
However, not all our observations were carried out under
photometric conditions or with the slit at the parallactic angle. 
The observed continuum range redward of the Ly$\alpha$
emission covered by the spectra ($1216(1+z)$ \AA\ to 9000 \AA) of these
high-redshift quasars is typically only few hundred \AA\ in the rest
frame. 
Moreover, some of the spectra have rather poor signal-to-noise ratio 
beyond 8000 \AA.
Thus, in some cases, it is difficult to determine accurately 
the continuum slope $\alpha$ from the observed spectrum.

However, the broad-band photometry gives us an alternative method to
determine the slope. 
The intrinsic quasar spectrum can be modeled as a power-law continuum
plus a series of emission lines. 
The equivalent width of the lines can be measured quite accurately
from the spectrum.
Therefore, the slope of the power-law can be determined by matching the
spectral model to the broad-band photometry. 
The SDSS $z'$ cutoff is beyond 10,000 \AA, so this gives a 
broader baseline over which to determine the slope.
Moreover, the absolute photometric calibration is more accurate.
The contribution to the broad-band colors from emission lines
other than Ly$\alpha$ is also small.
However, only bands which are not affected by Ly$\alpha$ forest lines
can be used in this method.
At $z>4.5$, Ly$\alpha$ moves into the $i'$ band, making it impossible
to determine the slope from the $z'$ measurement alone.
Therefore, we develop a method to fit the continuum slope $\alpha$ and
to derive $AB_{1450}$
of the quasar using information from both the observed spectrum and
the broad-band photometry.
Briefly, we construct a quasar spectral model that uses the observed
spectrum blueward of Ly$\alpha$ emission and the power-law continuum
plus emission line model redward of Ly$\alpha$ emission, and 
determine $\alpha$ and $AB_{1450}$ by finding the model that fits
the photometry the best. In detail, the method consists of following
steps:

 {\bf (1) Obtain a corrected observed spectrum $f_\lambda^{obs}$ that matches 
the SDSS photometry.}
For quasars with $i^* < 20$ and $z<4.3$, 
the photometric errors in the $r'$ and $i'$ bands
are small ($<0.04$ mag). We apply an additional  
correction function linear in wavelength
(with two free parameters, a zeropoint and a slope) to the spectrum so that
the synthetic photometry reproduces the SDSS $r^*$ and $i^*$ magnitudes.
For objects at $z>4.3$, the Ly$\alpha$ forest moves into $r'$, 
increasing the error in the $r'$ band, 
therefore we only apply a zeropoint shift to match the
$i^*$ magnitude. 

 {\bf (2) For a given $\alpha$ and $AB_{1450}$, construct a model spectrum
$f_\lambda^{model}$:}
\begin{equation}
f_\lambda^{model} = 
	\left\{ \begin{array}{ll}
		f_\lambda^{obs}& \mbox{if $\lambda <$ 1260 \AA $\times (1+z)$}, \\
		f_{1450} [\lambda/1450(1+z)]^{-\alpha-2} \cdot \sum_i [1+\mbox{EW}_i P(\lambda - \lambda_0^i)]& \mbox{if $\lambda >$ 1260 \AA $\times (1+z)$.}
		\end{array}
	\right.
\end{equation}

For $\lambda < 1260 \times (1+z)$, we use the flux from the corrected
observed spectrum $f_\lambda^{obs}$. 
The observed profile for the Ly$\alpha$+N V emission line 
is used as the line is strongly 
affected by the Ly$\alpha$ forest absorption.
For $\lambda > 1260 \times (1+z)$, 
the model spectrum has three components: 

(a) a power law continuum with $f_\lambda \propto
\lambda^{-\alpha-2}$ and continuum magnitude $AB_{1450}$;

(b) emission lines. 
We use the equivalent widths measured in Table 2 for
strong lines. For weak lines, we use  the line ratios in \cite{Francis91}
to scale from the measured Ly$\alpha$+N V emission line strength.
The line profile $P(\lambda)$ is assumed to be a Gaussian with FWHM
of 5000 km s$^{-1}$.
The choice of FWHM has negligible effect on the broad-band colors; 

(c) if there is any broad absorption or strong associated 
absorption system in the spectrum, the
equivalent widths and FWHMs are measured and the lines are
added to the spectrum. 

 {\bf (3) Obtain the best fit $\alpha$ and $AB_{1450}$.}
First, for a given $(\alpha, AB_{1450})$,
the synthetic SDSS magnitudes $m^{model}$ are calculated from the model spectrum.
We add the statistical error in Table 1 and the calibration error (\S 2)
of the SDSS photometry in quadrature to obtain the final photometric
error $\sigma^{obs}_i$. Then we minimize the total $\chi^2$ of the differences between the 
synthetic magnitudes and the SDSS photometry $m^{obs}$ with respect to
$\alpha$ and $AB_{1450}$:
\begin{equation}
\chi^2 = \sum_i \left( \frac{m^{model}_i - m^{obs}_i}{\sigma^{obs}_i} \right)^2.
\end{equation}
For quasars at $z<4.3$, $\lambda = 1260 \times (1+z)$ is still inside
the SDSS $r'$ band, while the $u'$ and $g'$ bands are dominated
by the absorption systems. Therefore,  
$r^*$, $i^*$ and $z^*$ magnitudes are used.
For quasars at $z>4.3$, only $i^*$ and $z^*$ magnitudes can provide
useful constraints on the continuum slope,  and 
only these bands are used in the fit.
The 1-$\sigma$ errors on the parameters are estimated
by finding the values that give $\chi^2 = \chi^2_{min} + 1$,
where $\chi^2_{min}$ is for the best-fit parameters.
The error on $\alpha$ is $\sim 0.20$ for $z<4.3$ and $\sim 0.5$ for
$z>4.3$, and the error on $AB_{1450}$ is typically a few percent.
The best-fit values and 1-$\sigma$ errors of $\alpha$ and $AB_{1450}$ 
are given in Table 3.
An example of the result of this fitting procedure is shown
in Figure 4. 

The major reason for determining $\alpha$ in this paper
is to estimate the selection completeness of the survey, as
the selection probability varies with different continuum shape (and therefore
different broad-band colors).
For objects with high signal-to-noise ratio spectra, the $\alpha$ values
determined from the spectra and from the method above are usually consistent
with each other within the quoted error.
Determining $\alpha$ using broad-band photometry does not depend on
the selection of a continuum window, and the error bar is easy to interpret.
The best-fit $\alpha$ using the method above always minimizes the difference
between model and observed colors.
By tying to the observed colors rather than fitting the observed spectra,
the effect of photometric calibration errors 
(at 2\% -- 7\% level in this paper) on the selection completeness calculation
is minimized:
an error on the photometric calibration or uncertainties 
in the filter response
would give rise to a wrong best-fit $\alpha$, 
but to first order,
the spectral model in \S 5 based on these $\alpha$ values 
should still reproduce the observed color distribution.

For the survey selection function (\S5) and the quasar luminosity function
(Paper IV), the calculations are carried out in two
cosmologies:
a model with $\Omega_m=1$ and $H_0=50$ km s$^{-1}$ km$^{-1}$,
where we refer to as the $\Omega=1$ model hereafter;
and a $\Lambda$-dominated flat model with 
$\Omega_m=0.35$, $\Lambda=0.65$ and $H_0=65$ km s$^{-1}$ km$^{-1}$,
which we refer to as the $\Lambda$-model hereafter.
In this paper, most of the results are presented using the $\Omega=1$ model.
Table 3 gives the absolute magnitude at 1450 \AA\ in the rest frame,
$M_{1450}$, for the $\Omega=1$ model.
In Paper IV, we use the $M_{1450}$  values to derive the quasar
luminosity function, and compare with low-redshift results after
converting to the rest-frame B-band. All the continuum properties of 
Table 3 have been corrected for Galactic extinction using the reddening
map of \cite{Schlegel98} assuming a standard extinction curve.
 
\subsection{Notes on Individual Objects}

\noindent {\bf SDSSp J012700.69--004559.1} ($z=4.06$).
A prominent damped Ly$\alpha$ absorption
line candidate is detected at $\lambda=5742$ \AA\ ($z_{abs}=3.72$), 
with a rest-frame equivalent width of 33 \AA.
The corresponding metal line systems (the C IV doublet at 7330/7317 \AA, 
and the Si IV doublet at 6632/6589 \AA) are also detected.

\noindent {\bf SDSSp J021043.17--001818.4} ($z=4.77$).
This is the highest redshift radio loud quasar known to date,
and is located only $10.4'$ away from SDSSp J021102.72--000910.3 ($z=4.90$).
The co-moving separation of the two quasars is 29.7 $h^{-1}$ Mpc for $\Omega=1$.
 
\noindent {\bf SDSSp J024457.19--010809.9} ($z=3.96$).
This is the previously-known quasar PSS 0244--0108, which has a reported
redshift of $z=4.01$ on the DPOSS website 
\footnote{\texttt{http://astro.caltech.edu/$^\sim$george/z4.qso}}.
It is a BAL quasar. Strong BAL troughs  due to Ly$\alpha$+N V and C IV are
clearly visible in the spectrum. The C IV BAL trough has a velocity of
$\sim 17200$ km s$^{-1}$ and FWHM of $\sim 8600$ km s$^{-1}$.
The redshift determination is not straightforward for BAL quasars due to
the presence of strong absorption features. The redshift reported
here is the average
of redshifts from the peaks of the O I and Si IV emission lines.

\noindent {\bf SDSSp J025019.78+004650.3} ($z=4.76$).
Typically, quasars at $z>4.5$ would encounter a Lyman Limit System within
0.1 of its emission line redshift (\cite{SSG91}, Paper I).
A strong Lyman Limit System (LLS) in this quasar is detected  only at $z=4.24$ 
($\lambda \sim 4780$ \AA). As a result, this object is detected at the 5-$\sigma$
level in the $g'$ band. If we assume the number density of the LLS
$N(z) = 0.27 (1+z)^{1.55}$ (\cite{Storrie94}), the fraction of
quasars at $z=4.76$ without a LLS at $z>4.24$ is 13\%.

\noindent {\bf SDSSp J030707.46--001601.4} ($z=3.70$).
A metal absorption system at $z=3.64$ is detected based on the C IV doublet 
at 7200/7186 \AA\ and the Si IV doublet at 6514/6471 \AA. 
A Mg II system at
$z=1.29$ is also detected (the doublet at 6430/6414 \AA).
 
\section{The Fall Equatorial Stripe Sample}

Using the selection criteria presented in \S 2 (Eq. 1 and 2), 
we define a complete color-selected
high-redshift quasar sample in the Fall Equatorial Stripe.
This complete sample covers the area of SDSS Runs 94 and 125. 
The right ascension of Run 94 ranges from 22$^h$22$^m$ to 3$^h$55$^m$. 
For Run 125, we only include areas with $23^h18^m < \alpha_{2000} < 4^h$, 
because the reddening increases to $E(B-V) > 0.2$ and becomes patchy
at $\alpha_{2000} > 4^h$  at the east end of Run 125.
After further correcting for (a) the overlap between Run 94 and 125 ($\sim$ 7\%),
(b) small areas of the sky with very poor seeing, 
and (c) areas on the CCDs affected by bright stars and defects which 
do not allow accurate measurements of the colors, the total effective
area of the sample is 181.9 deg$^2$.
During the selection process, we also rejected objects with 
flags indicating problems with image processing, in most cases associated
with deblending of close neighbors (see \cite{Lupton00}),
and some objects are further rejected upon visual inspection of the image.
The area covered by such objects is negligible.

The final color-selected sample contains 39 quasars at $3.6 < z < 5.0$. 
It includes 14 quasars presented in Paper I, 11 quasars presented in
\cite{HET2} and 14 quasars presented in this paper.
Figure 1 shows the positions of all the quasars presented in these
three papers on the color-color diagrams.
Several quasars in the same area of  the sky in these
papers are not part of the sample because they do not satisfy the final
color selection criteria, either due to the tighter selection 
compared to the criteria in Paper I, or due to slight changes
in the photometric zeropoints in the new photometric catalog.

In Figures 1(b) and (d), we plot the positions of 
the 14 non-quasars in the survey area
that satisfy the selection criteria in the color-color diagrams.
Among them, seven are galaxies with redshifts between 0.37 and 0.74.  
They all have compact morphology and are classified as stellar sources
in the photometric catalog. 
With the exception of one with poor signal-to-noise ratio, all are identified as
E+A (or k+A) galaxies
showing strong Balmer break, Balmer absorption lines and weak emission lines
(\cite{Dressler82}, \cite{Dressler99}).
Their broad-band colors are similar to those of high-redshift quasars:
the strong Balmer break mimics the Lyman break in the quasar spectrum,
while the continuum of E+A galaxies redward of the Balmer break is
also blue, due to the contributions of the A--F stars in the galaxy.
Five of the non-quasars are stars: one is a faint Carbon
star (Margon et al. 2000, in preparation), while the other four seem to be normal stars
scattered out of the stellar locus by photometric errors. 
We are not able to classify the other two objects.
One has  very poor signal-to-noise ratio;
we will discuss the nature of the other unclassified object in a separate paper
(Fan et al. 2000 in preparation).
Among the 53 objects that satisfy the final selection criteria, 
39 are quasars, giving a success rate of 
73\%. This is higher than the 63\% success rate quoted in Paper I.
Several of non-quasars in Paper I no longer satisfy the tighter final
selection with improved photometry. 
Note that most of the non-quasars are very close to the selection
boundary.

In Table 4, we present the properties of the 39 quasars.
The continuum magnitude and the continuum slope are calculated following
the procedures described in \S 3.1. For quasars in Paper I, these
quantities are re-measured based on improved (although not final) 
SDSS photometry. 
For the three previously known quasars for which we did not obtain new spectra,
we use the spectral properties
measured in \cite{Smith94} and \cite{K95} and the SDSS photometry
to derive their continuum magnitudes and power-law slopes.
Table 4 also lists the detection probability of each quasar, which
is described in the next section.

The sample in Table 4 covers a redshift range of $3.66 \leq z \leq 5.00$,
and an absolute magnitude range of $ -25.69 \leq M_{1450} \leq -27.72$
for the $\Omega=1$ model.
This is the largest complete sample of quasars at $z>3.6$ to date.
It includes 18 quasars at $z>4.0$ and six at $z>4.5$,
enabling us to determine the
quasar number density at $z \sim 4.8$ (Paper IV).
By comparison, the Schmidt, Schneider \& Gunn (1995) sample has 9 quasars
at $z>4.0$ with one at $z>4.5$, and the \cite{K95}  sample
includes 10 quasars at $z>4.0$ with a highest redshift of 4.45.

About 10\% of optically selected low-redshift quasars are BALs
(\cite{Weymann91}).
In this sample, two of the quasars 
(SDSSp J023909.98--002121.5 and SDSSp J024457.19--010809.9)
are classical BAL quasars; two other quasars
(SDSSp 015048.83+004126.2 and SDSSp 033910.53--003009.2) 
are candidate mini-BAL quasars.
The result is consistent with the fraction found in low-redshift samples.
However, some of the spectra have poor signal-to-noise ratio, and we
have not yet applied an objective criteria for selecting BALs to all the
spectra, so this fraction is likely to be a lower limit.
At least one object, SDSSp J033829.31+002156.3 ($z=5.00$), also shows
a strong and broad CIV absorption feature in a high signal-to-noise ratio
spectrum (\cite{Cowie99}).

Schmidt et al. (1995) and \cite{Stern00} found that about 10\% of known quasars at $z>4$ (most of
which are also color-selected) are detected
at  radio wavelengths at the mJy level.
Four quasars in our sample are FIRST radio sources with total flux of
4.0 -- 9.8 mJy at 20cm.
Although our sample size is still small, 
the fraction of radio-loud quasars in our sample
is consistent with the previous results.

\section{The Selection Function of the Color-Selected Sample}

To derive  the luminosity function of high-redshift quasars from
the color-selected sample in Table 4, 
we first need to calculate the selection function.
It is defined as 
the probability, $p(M_{1450}, z,  SED)$, that a quasar of a given
$M_{1450}$, $z$, and spectral energy distribution (SED) 
will satisfy the color selection criteria (Eqs. 1 and 2),  
In this section, we calculate the selection function 
using a Monte-Carlo simulation of quasar colors,
based on the quasar spectral model described in \cite{FanCss99}.
Similar procedures were adopted to calculate the  
selection function for multicolor
quasar surveys by \cite{Warren94} and \cite{K95}.

The measured colors of a quasar at a certain redshift and luminosity 
are determined by three factors:
its intrinsic SED; 
the absorption systems that are distributed randomly along the line of sight;
and the photometric errors.
In \S5.1, we describe the quasar spectral model and 
calculate the detection probability $p(M_{1450},z,  SED)$ 
of individual quasars,
taking into account the effects of the distribution of 
absorption systems and photometric errors.
We discuss in detail how this probability varies with different parameters,
and how the selection criteria affect the sample completeness.
In \S5.2, we calculate the distribution of the intrinsic properties of
quasars,
the continuum slopes $\alpha$ and equivalent widths of Ly$\alpha$+N V emission,
based on the sample in Table 4, after correcting for the detection
probability calculated in \S5.1.
Finally, in \S5.3, we calculate the average selection probability 
of the quasar sample as a function of redshift and luminosity.  

\subsection{Detection Probability of Individual Quasars}

To determine the detection probability, 
we first calculate the synthetic distribution of quasar colors
at a given $(M_{1450},z,  SED)$, following the procedures described in
\cite{FanCss99}. 
As in that paper and in \S3.1, 
the intrinsic quasar spectrum model, including
a power law continuum and a series of emission lines, 
is parameterized by the continuum slope $\alpha$ and the rest-frame
equivalent width of the Ly$\alpha$+NV emission line, EW(Ly$\alpha$+NV)
(abbreviated to EW in what follows).
The synthetic absorption spectrum takes into account 
intervening HI absorbers along the line of sight,
including Ly$\alpha$ forest systems, Lyman Limit Systems and damped
Lyman $\alpha$ systems.
We use the distribution functions of each kind of
absorber used by  \cite{FanCss99}.

We calculate the SDSS magnitudes from the model spectrum  
and add photometric errors in each band.  
Our precedures differ in two ways from those  of Appendix A of 
\cite{FanCss99}.
First, we use updated SDSS filter response functions 
as described in Appendix A.
They are slightly different from those of Fukugita et al. (1996);
in particular, the long-wavelength cutoff of the $g'$, $r'$, $i'$ and $z'$ 
filters are 100 -- 300 \AA\ bluer in the new measurements than
in the old ones.
For these broad-band filters, this difference has only a small effect
on the colors of normal stars.
However, it does affect the model quasar colors in certain redshift ranges 
for which the strong Ly$\alpha$ emission line is near
a filter edge.
Second, in the current calculation, we
use the measured seeing for Runs 94 and 125, $\sim 1.6''$,
to derive the photon noise of the PSF photometry.
The seeing in these commissioning runs is  worse than the $1.1''$ seeing
assumed in \cite{FanCss99}.
We further add the systematic calibration error in each band
in quadrature to the random errors, as in \S3.1.

For each object in the complete sample, 
a total of 1000 model quasars with the same $(M_{1450}, z, \alpha, EW)$
are generated, and SDSS magnitudes calculated.
The fraction of these which satisfy the selection criteria
of Eqs. (1) and (2) is defined to be the detection probability $p$,
as listed in Table 4.
It ranges from $\sim 100\%$ for
objects located in the part of the color-color diagrams 
farthest away from the stellar locus (see Figure 1) to 
$\sim 40\%$ for the objects near the boundary of the selection
criteria. 
An object with a given intrinsic spectrum can move in and out of the
selected region in color space due to differing amounts of 
absorption and photometric errors.
The average probability $\langle p \rangle = 0.80$ for the whole sample. 
This is considerably higher than that of Warren et al.
(1994), $\langle p \rangle \sim 0.5$; 
the higher photometric accuracy with CCDs
enables us to probe much closer to the stellar locus
in color space than do surveys using photographic plate data.

Note that the spectral model above does not include contributions
from BALs.
For the four possible BAL quasars in our sample, we
also calculate the detection probability by adding the observed BAL troughs
in the model spectra.
The detection probability changed by $\lesssim 10\%$ in all four cases.
There is no systematic effect and 
$\langle p \rangle$ remains at 0.78 for the whole sample.

The color-selection criteria in Eqs. (1) and (2) choose
regions in color space well-separated from the stellar locus.
Quasars with redder continua (larger $|\alpha|$) and weaker
emission lines are located in regions in the color-color diagrams
close to normal stars.
Any color-based selection will naturally select against these quasars. 
There may exist quasars with an intrinsic SED shape such that  
their broad-band colors are so similar to normal stars that their detection
probability is close to zero.
Thus it is difficult to correctly estimate the fraction 
of quasars that we would never find without assuming an {\em a priori} model
for the intrinsic distribution of $\alpha$ and EW.
The selection probability is also a strong function of the redshift
and luminosity of the quasar.
In Figure 5, we present the detection probability as a function
of $\alpha$ and EW at several different redshifts
and luminosities to
illustrate the dependence of $p$ on these parameters.

Figure 5(a) shows $p$ at $z=3.7$ and $M_{1450}=-27$ ($\Omega=1$ model);
this is near the lower redshift cutoff of the survey.  The quasar is
much brighter
than the flux limit, so photometric errors are not important.
In this case, a quasar with $\alpha \sim -0.8$ 
and EW = 70 \AA, which are the mean values we derive in \S5.2, 
has $p \sim$ 80\%; the completeness is rather high.
The completeness is lower for objects with redder continua 
or weaker emission lines.
The detectability drops to $\sim 20\%$ at $\alpha \sim -1.6$, 
where $i-z \gtrsim 0.25$.
This selection bias is introduced by the selection criterion
$i^* - z^* < 0.25$.
This eliminates a large number of compact
E+A galaxies at $z\sim 0.4$ which  have very similar $g^*-r^* $ and
$r^* - i^*$ colors to $z\sim 4$ quasars but are intrinsically redder
than are quasars. 
In the final selection of the SDSS main quasar sample, this condition
will be relaxed.
For objects with weak emission lines,
$p$ drops for bluer continua as well.
At $z\sim 3.7$, such objects could have $g^* - r^* < 1.0$,
and are thus excluded by the cut Eq. 1(c). 
This bias only affects redshifts close to the cutoff.

Figure 5(b) shows $p$ at $z=3.7$ and $M_{1450}=-25.5$,
corresponding to $m_{1450} \sim 20$.
This figure is for the same redshift as Figure 5(a), but
at the survey limiting luminosity.
In this case, $p$ is much lower.
At a given absolute magnitude, the {\em observed} apparent magnitudes
of a substantial fraction of quasars will be below the apparent magnitude cut
($i^* < 20.05$) due to photometric errors.
In addition, the large error in the $g'$ band
will scatter more quasars out of the selected region.
However, the selection is actually somewhat more sensitive to
larger $\alpha$ here than for brighter objects at the
same redshift (Figure 5a): $p \sim 20\%$ for $\alpha \sim -1.8$.
This is because the photometric error in $z'$ is $\sim 0.15$ mag,
scattering some objects with red intrinsic continua back 
into the sample with  $i^* - z^* \lesssim 0.25$.

At $z \sim 4.5$, the Ly$\alpha$ emission has moved out of the SDSS
$r'$ band, and lies in the gap between $r'$ and $i'$. 
Thus neither the $gri$ nor the $riz$ selection are
sensitive to these quasars.
Figure 5(c) shows the detection probability at this redshift 
and at $M_{1450} = -26.0$, near the survey limiting luminosity.
This is the region where the survey is most incomplete.
For typical values of $\alpha$ and emission line strength, $p \sim 50\%$.
At most redshifts, the presence of a strong Ly$\alpha$ emission line makes the
Lyman break in the spectrum more distinct, pushing the object
further away from the stellar locus in color space.
At $z\sim4.5$, however, with the Ly$\alpha$ emission line in the gap, 
quasars are closer to the stellar locus than either at higher or
lower redshift, making them difficult to select.

At $4.6 < z < 5.2$ (see Figure 1), the presence of strong Ly$\alpha$
forest absorption in the spectrum causes the object
to be very well separated from the stellar locus in the
$r^* - i^*$ vs. $i^* - z^*$ color-color diagram.
Figure 5(d) shows that at $z = 5.0$, the selection completeness
is very high, even for objects with weak emission lines or
red continua.
SDSS 1532--00 (\cite{BLLAC}),
a $z=4.62$ quasar without emission lines, was found in the SDSS
Spring Equatorial Stripe data using with the same selection
criteria. This  demonstrates that our selection is not sensitive to
emission line strength in this redshift range.

Thus, except in the redshift range $z\sim 4.4 - 4.5$ and
for luminosities near the survey limit, the detection probability
is high and the selection criteria are sensitive to a large range
of ($\alpha$, EW).
We will further examine the average selection function in \S5.3
after deriving the distribution functions of
the continuum slope and emission line strength in our sample.

\subsection{Distribution of Equivalent Width and Continuum Slope}

In this subsection, we use the sample in Table 4 to estimate the
distributions of $\alpha$ and EW(Ly$\alpha$+NV), 
after correcting the selection probability of each quasar.
We first calculate their weighted means, then
derive their distribution functions using a maximum likelihood 
approach and assume Gaussian distributions.
The histograms in Figure 6 show the observed distributions of 
the power-law continuum slope, $\alpha$,
and the rest-frame equivalent
width of Ly$\alpha$+NV emission, EW.  
Using only the 35 non-BAL quasars, and
weighting each object by the inverse of 
its selection probability $\frac{1}{p_i}$ (Table 4), 
we find a mean EW of 71.0 \AA\ with a standard deviation of 18.3 \AA,
at an average redshift $\langle z \rangle = 4.06$. 
Similarly, using all 39 quasars in the sample,
and applying a weight of $\frac{1}{p_i \delta_i^2(\alpha)}$, where
$\delta_i(\alpha)$ is the error on the $\alpha$ measurement,
we find a mean $\alpha$ of $-0.80$ with a standard deviation of 0.40.

However, these distributions can be biased if there is a population
of quasars with small $p$ which  are totally missing from the sample.
Thus, we use a maximum likelihood approach. 
The likelihood of the observed $(\alpha_i, \mbox{EW}_i)$ is given by:
\begin{equation}
L = \prod_i \frac{p_{i} f(\alpha_i) f(\mbox{EW}_i)}
{\int p(\alpha,\mbox{EW}) f(\alpha) f(\mbox{EW}) d\alpha\ d(\mbox{EW})} 
\end{equation}
where $f(\alpha)$ and $f(\mbox{EW})$ are the distribution functions of $\alpha$ and
EW. We assume both to be Gaussian, and uncorrelated with each other, 
\begin{equation}
f(\alpha_i) = \frac{1}{2\pi\sqrt{\delta_i^2+\sigma_\alpha^2}} 
\exp \left[ -\frac{(\alpha_i- \overline{\alpha} )^2}{2(\delta_i^2+\sigma_\alpha^2)} \right].
\end{equation}
where $\overline{\alpha}$ and $\sigma_\alpha$
are the mean and standard deviation of its intrinsic distribution.
Similarly, 
\begin{equation}
f(\mbox{EW}) = \frac{1}{2\pi\sigma_{\rm EW}}
\exp \left[ -\frac{(\mbox{EW}_i-\overline{\mbox{EW}})^2}{2\sigma_{\rm EW}^2} \right].
\end{equation}

Note that the normalizing integral $\int p(\alpha,\mbox{EW}) f(\alpha) f(\mbox{EW}) d\alpha\ d\mbox{EW}$ 
in the denominator of the likelihood function is essential;
without it, the maximum likelihood solution would not depend on
the choice of $p$, which is clearly wrong.
However, as we show in \S5.1, 
the selection function $p$ is a complicated function
of $M_{1450}$, $z$, $\alpha$ and EW.
This integral cannot be evaluated without knowing 
the underlying distribution function of $z$ and $M_{1450}$, namely, the
luminosity function.
A full solution requires maximizing the likelihood of the full distribution
function $p(M_{1450}, z, \alpha, \mbox{EW}) f(M_{1450}, z, \alpha, \mbox{EW})$. 
We will present the maximum likelihood solution of the luminosity function
in Paper IV,
but we find that the distributions of $\alpha$ and EW are rather
insensitive to the choice of luminosity function,
as it provides a weighting of the probability function, which is 
only a slow varying function of $z$ and $M_{1450}$.
Using the luminosity function we find in Paper IV (for the $\Omega=1$ model), 
\begin{equation}
\log \Phi(M_{1450}, z) = -7.24 - 0.48 (z-3) + 0.63 (M_{1450} + 26),
\end{equation}
where $\Phi(M_{1450}, z)$ is the cumulative luminosity function
in units of comoving Mpc$^{-3}$, we find:
$\rm \langle EW \rangle = 69.3$ \AA, $\rm \sigma_{EW} = 18.0$ \AA, 
$ \langle \alpha \rangle = -0.79 $, and $\sigma_{\alpha} = 0.34$.
In Figure 6, the average $\alpha$ value from the maximum likelihood
solution is clearly
offset from the {\em observed} distribution, due to the selection
bias against objects with redder continua.
The results above assume that the distributions of
$\alpha$ and EW are not correlated with each other or with
redshift or luminosity.
Luminosity and emission line strength are known to be correlated
(the Baldwin (1977) effect),
although for the small luminosity range (2 mag) our sample covers,
this is a rather small effect.
These results change by less than 5\% for a change of the parameters in
the assumed luminosity function by $\sim 30\%$.

These statistics can be compared with results from previous studies.
Based on 30 quasars at $3.1 < z < 4.75$, Schneider et al. (1991)  found
$ \rm EW(Ly\alpha+NV) = 82.3 \pm 29.8$ \AA.
Note that most of the quasars in Schneider et al. (1991)
are selected based on their slitless spectra, and are biased
to quasars with stronger emission lines.
\cite{SSB} found $ \rm EW(Ly\alpha+NV) = 80.9 \pm 29.8$ \AA\
for a sample of 59 quasars at $z>2.75$.
\cite{Warren94} found a median equivalent width of
$ \rm EW(Ly\alpha+NV) = 68 $ \AA\ for quasars at $z>2.2$.
Our result is consistent with these high-redshift samples.
At lower redshift, the Large Bright Quasar Survey (LBQS) sample
has EW(Ly$\alpha$+NV) = 52 \AA\ (Francis et al. 1991).

Schneider et al. (1991) found  $\alpha = -0.92 \pm 0.26$, and \cite{SSB}
found $\alpha = -0.78 \pm 0.27$, both consistent with our result.
Our value is redder than the values found from low redshift samples
(e.g., the LBQS sample shows $\alpha_{med} = -0.3$).
This could be due to the fact that the $\alpha$ values from
low and high redshift samples are based on different wavelength
ranges. 
A detailed comparison of the continuum slope for different
redshift ranges would require observations of high-redshift quasars
at IR wavelengths.
However, our purpose in determining $\alpha$ is to reproduce
the true color distribution of the underlying quasar population
and thus to calculate the selection function.
Since we measure $\alpha$ by matching to the SDSS photometry,
any bias will be small because of the limited wavelength range
in the optical spectrum of high-redshift quasars.
The difference in the continuum slope between the low and high redshift
sample and any deviations of the underlying continuum from a power law, 
however, will affect the calculation of the K-correction, and
the evolution of the luminosity function  from $z \lesssim 3$ to
$z > 4$ (Paper IV).

\subsection{Selection Completeness of the Quasar Sample}
 
In Figure 7, we present the average selection probability for $M_{1450} = -26.5$, 
as functions of $\alpha$ and EW, $\langle p(\alpha) \rangle$,
and $\langle p(\mbox{EW}) \rangle$, 
averaged over the survey redshift range $3.6 < z < 5.0$.
$\langle p(\alpha) \rangle$ is defined as:
\begin{equation}
\langle p(\alpha) \rangle = \frac{\int_{3.6}^{5.0} p(M_{1450}, z, \alpha)
f(\mbox{EW}) \Psi(M_{1450}, z) \frac{dV(z)}{dz} dz}
{\int_{3.6}^{5.0} f(\mbox{EW}) \Psi(M_{1450}, z) \frac{dV(z)}{dz} dz}
\end{equation}
where $\Psi(M_{1450}, z) = d\Phi(M_{1450}, z)/dM$ 
is the differential luminosity
function. The quantity $\langle p(\mbox{EW}) \rangle$ is similarly defined.
This figure illustrates the selection completeness as a function of
the quasar SED in \S5.1 more clearly.
On average, the selection criteria are more sensitive to quasars with
bluer continua. At this luminosity, where the effects of photometric errors
are small, the average selection probability $p \gtrsim 80\%$ for
$\alpha > -0.9$, but drops to $\sim 40\%$ for $\alpha=-1.5$, and to $\sim 20\%$
for $\alpha=-2.0$.
Note that $\alpha=-1.5$ is about 2 $\sigma$ away from the mean $\alpha$.
Thus although the current survey is less complete for redder quasars,
those objects are not excluded completely and can be corrected for by
applying the selection function.
The figure also shows that the average selection probability is higher
than 80\% for quasars with EW $>$ 40 \AA, and
is still higher than 50\% for quasars with much weaker lines,
EW $\sim$ 20 \AA.

Finally, we calculate the average selection probability of 
the high-redshift quasars in our sample
as a function of redshift and luminosity $p(M_{1450}, z)$,
averaging over the $\alpha$ and EW distributions.
We calculate the quasar model colors following the same procedures
described in \S5.1, and assume that the intrinsic properties of quasars
follow the distributions we obtained in \S5.2.
The selection function is calculated for both $\Omega=1$ and $\Lambda$ models
for $-25 > M_{1450} > -30$ and $3.6 < z < 5.2$.

The results of the selection function are summarized in Figure 8
for both cosmologies.
The heavy solid line in the figure represents the 5\% contour, 
indicating the survey limiting absolute magnitude at each redshift. 
Our survey is cut at $i^* < 20.05$
for $gri$ selected quasars, and at $i^*<20.2$ for $riz$ selected quasars.
For the $\Omega=1$ model, the survey reaches $M_{1450}\sim -25.3$ 
at the lowest redshift $z\sim 3.6$,
and to $M_{1450} \sim -25.8$ at $z \sim 4.5$.
The survey then goes slightly deeper at $4.5 < z < 5.0$, 
due to the fact that our $riz$ selection
goes slightly deeper than does the $gri$ selection. 
Not surprisingly, the survey is more complete for 
brighter quasars due to smaller photometric errors.
The figure indicates that over the entire redshift range between 3.6 and 5.2,
the survey is complete to at least 40\% for quasars more than
$\sim 0.4$ mag brighter
than the survey limit.
The main difference between the two cosmologies is a slight shift in the 
$M_{1450}$ axis.

The survey completeness is a function of redshift:
it is most complete in the redshift ranges of 3.8 -- 4.2 and
4.7 -- 5.2, with $p > 0.8$ for bright quasars. 
As mentioned in \S5.1, the lower completeness 
at $z\sim3.6$ is due to the $g^*-r^* > 1.0$ cut in
the $gri$ selection. 
The completeness is lowest at $z\sim4.4 - 4.5$,
about 50\% for bright quasars. 
This is the region in which the Ly$\alpha$ emission moves between 
the $r'$ and  $i'$ filters, 
and the quasar moves from being selected with the $gri$ selection of Eq. (1)
to being selected with the  $riz$ selection of Eq. (2).
The final quasar target selection algorithm in the SDSS uses a 
selection in multi-dimensional color space, and is expected to be
more complete than the results presented here (see, e.g., \cite{Newberg97}).

The locations of the 39 quasars in the sample are also plotted on
Figure 8. 
Note that the average detection probability in Figure 8 is different
from the values in Table 4, which is the detection probability
for a specific type of intrinsic spectrum.
Using the contours in Figure 8, we find an overall average 
$\langle p(M_{1450},z) \rangle = 0.75$ for the 39 quasars,
which is slightly lower than $\langle p(M_{1450}, z, \alpha, \mbox{EW}) \rangle = 0.80$ 
we found in \S5.1. 
 
\section{Summary}

 In this paper, we present a color-selected flux-limited sample of
high-redshift quasars in the Fall Equatorial Stripe that includes
39 quasars at $3.6 < z < 5.0$ and $i^* \lesssim 20$, covering
an effective area of 182 deg$^2$.
These objects are identified from 53 quasar candidates, giving
a success rate of 73\%.

 From this sample, we estimate the average power law slope
of the quasar spectrum in the rest-frame UV to be $-0.79$ with
a standard deviation of 0.34; and the rest-frame equivalent width
of the Ly$\alpha$+NV emission line to be 69 \AA\ with a standard deviation
of 18 \AA. These results are consistent with previous high-redshift
quasar samples.

 We derive the selection function of the survey as a function
of redshift, luminosity and the quasar SED shape.
We show that on average, the selection criteria are more sensitive
to quasars with bluer continua; the survey is incomplete for
red quasars with $\alpha < -1.8$.
The selection probability does not strongly depend on the quasar
emission line strength.
The selection completeness is also a function of redshift.
The survey is the most complete in the redshift range
from 3.8 to 4.2 and 4.7 -- 5.2, 
while it suffers $\sim 50\%$ incompleteness at $z\sim 4.5$.
The average selection probability for the whole sample is $\sim 75\%$.
In Paper IV, we use the high-redshift quasar sample and the
survey selection function presented in this paper
to study the evolution of the quasar luminosity function.

\bigskip
The Sloan Digital Sky Survey (SDSS) is a joint project of the
University of Chicago, Fermilab, the Institute for Advanced Study, the
Japan Participation Group, The Johns Hopkins University, the
Max-Planck-Institute for Astronomy, Princeton University, the United
States Naval Observatory, and the University of Washington.  Apache
Point Observatory, site of the SDSS, is operated by the Astrophysical
Research Consortium.  Funding for the project has been provided by the
Alfred P. Sloan Foundation, the SDSS member institutions, the National
Aeronautics and Space Administration, the
National Science Foundation, the U.S. Department of Energy, and 
Monbusho, Japan.
The SDSS Web site is {\tt http://www.sdss.org/}.
The Hobby-Eberly Telescope (HET) is a joint project of the University of Texas
at Austin,
the Pennsylvania State University,  Stanford University,
Ludwig-Maximillians-Universit\"at M\"unchen, and Georg-August-Universit\"at
G\"ottingen.  The HET is named in honor of its principal benefactors,
William P. Hobby and Robert E. Eberly.  
XF and MAS acknowledge 
support from Research Corporation, NSF grant AST96-16901, the
Princeton University Research Board, and a Porter O. Jacobus Fellowship.
WHB acknowledges support from
the Institute of Geophysics and Planetary Physics (operated under the
auspices of the U.S. Department of Energy by the University of California
Lawrence Livermore National Laboratory under contract No.~W-7405-Eng-48).
DPS acknowledges support from NSF grant  AST99-00703.

\appendix
\section{Updated SDSS Photometric System Response Function}

In Spring 2000, the SDSS photometric system 
response functions of the photometric
camera on the 2.5m telescope were re-measured using a monochrometer. 
It was found that the passbands of the $g'$, $r'$, $i'$ and $z'$
filters had changed since the filters were measured prior to their
installation in the camera in 1997.  
The major effect is that the red edges of these passbands are at 
wavelengths 3\% bluer 
than the designed values (\cite{F96}).
For the $g'$, $r'$ and $i'$ filters,
the red cutoff is accomplished by an interference coating, whereas the blue
cutoff is done by a long-pass colloidal gold-in-glass glass filter.  
In the SDSS 2.5m imaging camera, the filters are cemented on
the  back face of the corrector, and the interference films are
exposed to the dewar vacuum (Gunn et al. 1998).
The most plausible explanation for the change is that the oxide layers in the
interference films have dehydrated in the vacuum 
and the refractive index of the layers have slightly changed.
The manufacturer, Asahi-Spectra Ltd, Tokyo, confirms that this is
likely. 
The change in the $z'$ response is due to the improved measurement
of the quantum efficiency at long wavelength for the thick CCDs used
in the camera.

Figure A.1 presents the filter response curves for the new and old measurements.
Table A.1 gives the effective wavelengths (defined in Eq. 3 in \cite{F96})
and FWHMs of the SDSS filters of the SDSS 2.5m imaging
camera according to the new measurements.
Although these new measurements are still preliminary, 
calculations are carried out with the new system
response curve throughout this paper.

\section{Faint Quasars from the Fall Equatorial Stripe}

The SDSS will image the Fall Equatorial Stripe 35 -- 40 times
during the five-year survey. Since the  Fall Equatorial Stripe
is in the southern Galactic Cap, it is called the SDSS Southern Survey
(\cite{York00}). 
The final combined images of the Southern Survey will reach $r' \sim 25$,
2 magnitudes deeper than the main survey. 
These data will be used to search for fainter quasars,  and to study the 
evolution and large scale distribution of quasars at much lower luminosities.
During the commissioning phase of the survey, part of the  Fall Equatorial Stripe
has been imaged twice. The combined catalog from the two observations
is used to select quasar candidates at $i^* < 21$, where the
photometric error from  a single observation is too large 
for efficient candidate selection.
Eighteen high-redshift quasars are identified from these candidates.
The algorithm to combine the imaging data and to search for faint quasar
candidates is still under development, and the sample is far from complete. 
In this Appendix, we simply present the SDSS photometry and the spectra of
these 18 quasars, some of which exhibit interesting spectral features.

Table B.1 lists the SDSS imaging runs we used in the selection
of faint quasars.
The overlapping regions of Runs 94/259 (the Northern Strip of the
Stripe, see York et al. 2000) and
Runs 125/273 (Southern Strip) were used.
Tables B.2, B.3 and B.4,  similar to Tables 1, 2 and 3, present the
photometry, emission line, and continuum properties of these 18 quasars.
In Table B.2, the photometry in both observations are listed;
in Table B.4, we do not measure the continuum slope from the photometry, 
due to the increasing photometric errors, especially in the $z'$ band.
Figure B.1 gives the finding charts for these quasars.
Figure B.2 presents the spectra. Sixteen of the spectra were obtained
with the Keck II/LRIS during the 1999 October  observing run.
Three of the faint quasars have very poor signal-to-noise ratio beyond
8000 \AA\, thus we do not show this spectral region in the figure.
Two objects (SDSSp J024434.87+000124.9 and SDSSp J024452.33--003318.0) were
observed with the ARC 3.5m telescope in 1999 December and 2000 January.

\subsection{Note for Individual Objects}

\noindent
{\bf SDSSp J010905.81+001617.1} ($z=3.68$).
This has a number of absorption systems in the spectrum, identified
both by their C IV $\lambda$1548,1551 
and Si IV $\lambda$1394,1403 absorption lines, including two narrow-line 
systems,  at $z=3.71$ (higher than the emission-line redshift), 
and at $z=3.59$, respectively; and two broad features at $z=3.43$ (with C IV absorption centered
at 6855 \AA\ and Si IV at 6233 \AA), and at $z=3.33$ (with C IV and Si IV absorption
at 6700 \AA\ and 6063 \AA).

\noindent
{\bf SDSSp J015015.58+004555.7} ($z=3.91$).
This is a BAL quasar, with a velocity (average of C IV and Si IV features)
of $\sim 7300$ km s$^{-1}$.

\noindent
{\bf SDSSp J023749.33+005715.6} ($z=3.57$).
This quasar has very narrow emission lines. 
The Ly$\alpha$ and N V emission lines are
clearly separated. The Si IV emission line  shows
two peaks, from the Si IV $\lambda$1394, 1403 components.
The average FWHM of the emission lines is only $\sim 1500$ km s$^{-1}$.
Quasars (and Seyfert 1s) with FWHM less than 2000 km s$^{-1}$ are
usually classified as Narrow Line Quasars (or Narrow Line Seyfert 1s,
which typically have lines broader than Seyfert 2s and with smaller
[O III]/H$\beta$ ratio, see, e.g. \cite{Baldwin88}, \cite{Osterbrock85}).
Those object tend to have strong Fe II emission
and soft X-ray spectra, and are thought to be either AGNs with face-on
orientation, or represent a higher accretion rate (e.g., \cite{Leighly99}
and reference therein).
This is the first narrow line quasar at $z>3.5$ of which we are aware.
A damped Ly$\alpha$ candidate is also detected at $z=3.26$ ($\lambda$ = 5177 \AA).

\noindent
{\bf SDSSp J024347.37--010611.7} ($z=3.90$).
The strong absorption at 7268/7286 \AA\ is due to a Mg II absorption system
at $z=1.600$. The total rest-frame equivalent width is 12.3 \AA.
Absorption lines corresponding to several FeII features at the same redshift
are also visible (at 6758\AA, 6723\AA, 6193\AA, 6172\AA\ and 6094 \AA).

\noindent
{\bf SDSSp J031427.92+002339.4} ($z=3.68$).
Several metal line systems are detected: two C IV/Si IV systems are at $z=3.55$
 and $z=3.48$,  and there is a Mg II system at $z=1.11$.

\noindent
{\bf SDSSp J032459.10--005705.1} ($z=4.80$).
This is a very odd-looking high-redshift quasar. 
The Ly$\alpha$+N V emission is clearly affected by strong absorption,
and the emission line seems very narrow. The O I emission is
barely detected, but it is affected by the telluric A-band. 
The redshift is determined by a fit to the O I line.

\noindent
{\bf SDSSp J033505.43+010337.2} ($z=3.58$).
Most of the rich absorption lines in the spectrum of this quasar seem to be actually
due to one very strong associated absorption system at the same redshift
as the emission lines.
A entire suite of absorption features due to different ions can be  identified,
including N V $\lambda$1238+1240, Si II $\lambda$1260, O I $\lambda$1302, Si II $\lambda$1304,
C II $\lambda$1334, Si IV $\lambda$1393+1402, Si II $\lambda$1526 and
C IV $\lambda$1548+1550.

\begin{scriptsize}
\begin{deluxetable}{cccccccr}
\tablenum{1}
\tablecolumns{8}
\tablewidth{0pc}
\tablecaption{Positions and Photometry of Bright SDSS High-redshift Quasars}
\tablehead
{
SDSS name & redshift & $u^*$ & $g^*$ & $r^*$ & $i^*$ & $z^*$ & run
}
\startdata
SDSSp J001950.06$-$004040.9 & 4.32 $\pm$ 0.01 & 23.60 $\pm$ 0.86 & 21.11 $\pm$ 0.05 & 19.76 $\pm$ 0.02 & 19.61 $\pm$ 0.03 & 19.59 $\pm$ 0.10 &  125 \\ 
SDSSp J005922.65$+$000301.4$^a$ & 4.16 $\pm$ 0.01 & 23.68 $\pm$ 0.77 & 22.47 $\pm$ 0.16 & 19.25 $\pm$ 0.01 & 19.18 $\pm$ 0.02 & 19.06 $\pm$ 0.07 &  125 \\
SDSSp J010822.70$+$001147.9 & 3.71 $\pm$ 0.01 & 23.25 $\pm$ 0.69 & 20.67 $\pm$ 0.03 & 19.49 $\pm$ 0.04 & 19.44 $\pm$ 0.03 & 19.57 $\pm$ 0.11 &  125 \\ 
SDSSp J012019.99$+$000735.5 & 4.08 $\pm$ 0.01 & 24.73 $\pm$ 0.61 & 21.47 $\pm$ 0.06 & 20.03 $\pm$ 0.03 & 19.84 $\pm$ 0.03 & 20.14 $\pm$ 0.17 &  125 \\ 
SDSSp J012700.69$-$004559.1 & 4.06 $\pm$ 0.01 & 22.99 $\pm$ 0.63 & 19.76 $\pm$ 0.02 & 18.37 $\pm$ 0.01 & 18.10 $\pm$ 0.01 & 18.08 $\pm$ 0.02 &  125 \\ 
SDSSp J013108.19$+$005248.2 & 4.19 $\pm$ 0.01 & 22.88 $\pm$ 0.51 & 22.05 $\pm$ 0.10 & 20.40 $\pm$ 0.04 & 20.05 $\pm$ 0.04 & 20.26 $\pm$ 0.15 &  125 \\
SDSSp J020427.81$-$011239.6 & 3.91 $\pm$ 0.01 & 22.53 $\pm$ 0.31 & 20.97 $\pm$ 0.04 & 19.74 $\pm$ 0.02 & 19.65 $\pm$ 0.02 & 19.49 $\pm$ 0.07 &  125 \\ 
SDSSp J021043.17$-$001818.4 & 4.77 $\pm$ 0.02 & 23.77 $\pm$ 0.69 & 22.84 $\pm$ 0.17 & 20.66 $\pm$ 0.04 & 19.25 $\pm$ 0.02 & 19.27 $\pm$ 0.06 &  125 \\
SDSSp J024457.19$-$010809.9$^b$ & 3.96 $\pm$ 0.01 & 22.72 $\pm$ 0.35 & 20.00 $\pm$ 0.02 & 18.60 $\pm$ 0.01 & 18.38 $\pm$ 0.01 & 18.18 $\pm$ 0.02 &  125 \\ 
SDSSp J025019.78$+$004650.3 & 4.76 $\pm$ 0.01 & 23.90 $\pm$ 0.65 & 22.87 $\pm$ 0.17 & 21.06 $\pm$ 0.05 & 19.67 $\pm$ 0.03 & 19.56 $\pm$ 0.09 &  94 \\
SDSSp J030025.23$+$003224.2 & 4.19 $\pm$ 0.01 & 22.74 $\pm$ 0.87 & 21.96 $\pm$ 0.09 & 19.97 $\pm$ 0.02 & 19.96 $\pm$ 0.03 & 19.75 $\pm$ 0.08 &  125 \\ 
SDSSp J030707.46$-$001601.4 & 3.70 $\pm$ 0.01 & 22.75 $\pm$ 0.46 & 21.24 $\pm$ 0.05 & 20.03 $\pm$ 0.02 & 19.95 $\pm$ 0.03 & 20.07 $\pm$ 0.13 &  125 \\ 
SDSSp J035214.33$-$001941.1 & 4.18 $\pm$ 0.01 & 23.83 $\pm$ 0.72 & 21.73 $\pm$ 0.08 & 20.11 $\pm$ 0.02 & 19.73 $\pm$ 0.03 & 19.58 $\pm$ 0.09 &  125 \\ 
SDSSp J040550.26$+$005931.2 & 4.05 $\pm$ 0.01 & 22.59 $\pm$ 0.46 & 22.30 $\pm$ 0.12 & 20.22 $\pm$ 0.03 & 20.01 $\pm$ 0.04 & 19.65 $\pm$ 0.09 &  125 \\ 
SDSSp J232717.99$+$000546.1 & 3.66 $\pm$ 0.01 & 24.67 $\pm$ 0.77 & 20.92 $\pm$ 0.04 & 19.89 $\pm$ 0.03 & 19.85 $\pm$ 0.04 & 19.95 $\pm$ 0.16 &  125 \\ 
SDSSp J235053.55$-$004810.3 & 3.85 $\pm$ 0.01 & 23.60 $\pm$ 0.86 & 21.11 $\pm$ 0.05 & 19.76 $\pm$ 0.02 & 19.61 $\pm$ 0.03 & 19.59 $\pm$ 0.10 &  125 \\ 
\enddata
\tablenotetext{}{Positions are in J2000.0 coordinates;
asinh magnitudes (Lupton, Gunn \& Szalay 1999) are
quoted; errors are statistical only.  For reference, zero flux
corresponds to asinh magnitudes of 23.40, 24.22, 23.98, 23.51, and
21.83 in $u^*, g^*, r^*, i^*$, and $z^*$, respectively.}
\tablenotetext{a}{This is the previously known quasar PSS 0059+0003 (Kennefick et al. 1995)}
\tablenotetext{b}{This quasar was first reported at http://astro.caltech.edu/$^{\sim}$george/z4.qsos as PSS 0244-0108}
\end{deluxetable}
\end{scriptsize}

\begin{scriptsize}
\begin{deluxetable}{cccccc}
\tablenum{2}
\tablecolumns{6}
\tablewidth{0pc}
\tablecaption{Emission Line Properties of Bright SDSS High$-$redshift Quasars}
\tablehead
{
quasar           &      O$\,$VI 1034  & Ly$\alpha$ 1216 $+$ 1240 & O$\,$I+Si$\,$II 1306 & Si$\,$IV+O$\,$IV]  1402&
 C$\,$IV 1549 
}
\startdata
SDSSp J001950.06$-$004040.9 &  ... & 6637.1 $\pm$ 4.0 & 7443.9 $\pm$ 7.4 & 6948.3 $\pm$ 6.9 & 8247 $\pm$ 7.7\\ 
 &  ... &  53.5 $\pm$  0.8 &   1.4 $\pm$  0.6 &   3.3 $\pm$  0.7 &  13.4 $\pm$  2.6\\ 
SDSSp J010822.70$+$001147.9 &  ... & 5728.2 $\pm$ 0.1 & 6151.2 $\pm$ 2.3 & 6572.2 $\pm$ 2.0 & 7281 $\pm$ 0.8\\ 
 &  ... &  74.6 $\pm$  0.6 &   2.1 $\pm$  0.3 &   6.6 $\pm$  0.4 &  30.3 $\pm$  0.8\\ 
SDSSp J012019.99$+$000735.5 & 5275.5 $\pm$ 0.6 & 6191.2 $\pm$ 0.5 & 6647.5 $\pm$ 2.4 & 711.4 $\pm$ 3.1 & 7868 $\pm$ 1.1\\ 
 &  23.0 $\pm$  0.6 &  82.9 $\pm$  1.0 &   3.8 $\pm$  0.5 &   7.3 $\pm$  0.9 &  31.6 $\pm$  1.0\\ 
SDSSp J012700.69$-$004559.1 & 5226.5 $\pm$ 0.2 & 6194.1 $\pm$ 0.1 &  ... & 7192.5 $\pm$ 0.5 & 7840 $\pm$ 0.3\\ 
 &  14.9 $\pm$  0.1 &  93.5 $\pm$  0.1 &  ... &  14.1 $\pm$  0.2 &  33.2 $\pm$  0.2\\ 
SDSSp J013108.19$+$005248.2 & 5369.5 $\pm$ 1.1 & 6320.2 $\pm$ 0.2 & 6779.1 $\pm$ 2.5 & 7266.7 $\pm$ 3.3 & 8046 $\pm$ 0.9\\
 &  19.1 $\pm$  0.9 &  71.6 $\pm$  1.8 &   5.1 $\pm$  0.7 &   7.5 $\pm$  0.8 & 25.0 $\pm$  0.9\\
SDSSp J020427.81$-$011239.6 &  ... & 6007.1 $\pm$ 0.5 & 6435.5 $\pm$ 2.0 & 6893.8 $\pm$ 2.4 & 7592 $\pm$ 1.8\\ 
 &  ... &  71.1 $\pm$  0.5 &   4.2 $\pm$  0.4 &  11.6 $\pm$  0.7 &  25.0 $\pm$  1.2\\ 
SDSSp J021043.17$-$001818.4 &  ... & 7019.6 $\pm$ 0.8 &  ... &  ... &  ... \\
 &  ... &  42.5 $\pm$  0.7 &  ... &  ... &  ... \\
SDSSp J024457.19$-$010809.9 &  ... & 6121.1 $\pm$ 0.8 &  ... & 6950.8 $\pm$ 6.5 & 7691 $\pm$ 2.8\\ 
 &  ... &  87.9 $\pm$  0.9 &  ... &  11.0 $\pm$  0.9 &  36.7 $\pm$  1.1\\ 
SDSSp J025019.78$+$004650.3 &  ... & 7034.1 $\pm$ 0.7 &  ... & 8068.6 $\pm$ 3.0 & ... \\
 &  ... &  42.8 $\pm$  1.0 &  ... &   4.1 $\pm$  0.4 &  ... \\
SDSSp J030025.23$+$003224.2 & 5350.0 $\pm$ 0.2 & 6309.2 $\pm$ 0.1 & 6774.2 $\pm$ 2.5 & 7259.2 $\pm$ 3.0 & 8038 $\pm$ 0.4\\
 &  16.4 $\pm$  0.2 & 114.7 $\pm$  0.7 &   2.4 $\pm$  0.3 &   9.2 $\pm$  0.7 & 33.1 $\pm$  0.4\\
SDSSp J030707.46$-$001601.4 &  ... & 5717.5 $\pm$ 0.2 & 6135.9 $\pm$ 0.9 & 6586.9 $\pm$ 1.4 & 7270 $\pm$ 0.7\\
 &  ... &  60.3 $\pm$  0.7 &   5.6 $\pm$  0.2 &  11.0 $\pm$  0.5 &  13.7 $\pm$ 0.2\\
SDSSp J035214.33$-$001941.1 &  ... & 6321.9 $\pm$ 0.4 & 6783.3 $\pm$ 3.7 & 7256.5 $\pm$ 2.6 & 8015 $\pm$ 2.3\\
 &  ... & 108.3 $\pm$  1.6 &   1.9 $\pm$  0.6 &   8.3 $\pm$  0.8 &  17.9 $\pm$ 0.1\\
SDSSp J040550.26$+$005931.2 &  ... & 6144.5 $\pm$ 0.1 & 6593.8 $\pm$ 2.1 & 7073.0 $\pm$ 2.0 & 7826 $\pm$ 0.5\\ 
 &  ... &  95.3 $\pm$  1.2 &   3.8 $\pm$  0.4 &   6.0 $\pm$  0.4 &  30.3 $\pm$  0.4\\ 
SDSSp J232717.99$+$000546.1 &  ... & 5687.0 $\pm$ 0.9 &  ... & 6538.0 $\pm$ 3.6 & 7219 $\pm$ 1.1\\ 
 &  ... &  50.6 $\pm$  0.7 &  ... &   7.4 $\pm$  0.9 &  18.9 $\pm$  0.7\\ 
SDSSp J235053.55$-$004810.3 &  ... & 5934.4 $\pm$ 0.7 & 6334.1 $\pm$ 3.3 & 6810.4 $\pm$ 3.6 & 7501 $\pm$ 2.3\\ 
 &  ... &  52.8 $\pm$  0.6 &   9.3 $\pm$  2.1 &   7.8 $\pm$  0.8 &  15.1 $\pm$  0.6\\ 
\enddata
\tablenotetext{}{The two entries in each line are the central wavelength and rest frame equivalent width
from the Gaussian fit to the line profile, both measured in
\AA{}ngstroms.}
\end{deluxetable}
\end{scriptsize}

\newpage

\begin{deluxetable}{cccccr}
\tablenum{3}
\tablecolumns{6}
\tablecaption{Continuum Properties of Bright SDSS High$-$redshift Quasars}
\tablehead{quasar       & redshift & E(B$-$V) & AB$_{1450}$ &$M_{1450}$ &$\alpha
$ }
\startdata
SDSSp J001950.06$-$004040.9 & 4.32 $\pm$ 0.01 & 0.027 & 19.62 $\pm$ 0.05 & $-26.36$ &  $-0.02$ $\pm$ 0.62\\
SDSSp J005922.65$+$000301.4 & 4.16 $\pm$ 0.01 & 0.025 & 19.30 $\pm$ 0.03 & $-26.62$ &  $-1.09$ $\pm$ 0.33 \\
SDSSp J010822.70$+$001147.9 & 3.71 $\pm$ 0.01 & 0.029 & 19.62 $\pm$ 0.04 & $-26.12$ &  $-0.19$ $\pm$ 0.24\\
SDSSp J012019.99$+$000735.5 & 4.08 $\pm$ 0.01 & 0.037 & 19.96 $\pm$ 0.04 & $-25.93$ &  $-0.52$ $\pm$ 0.30 \\
SDSSp J012700.69$-$004559.1 & 4.06 $\pm$ 0.01 & 0.031 & 18.28 $\pm$ 0.02 & $-27.60$ &  $-0.66$ $\pm$ 0.13 \\
SDSSp J013108.19$+$005248.2 & 4.19 $\pm$ 0.01 & 0.022 & 20.17 $\pm$ 0.04 & $-25.76$ & $0.11$ $\pm$ 0.34 \\
SDSSp J020427.81$-$011239.6 & 3.91 $\pm$ 0.01 & 0.026 & 19.80 $\pm$ 0.03 & $-26.02$ &  $-0.83$ $\pm$ 0.17 \\
SDSSp J021043.17$-$001818.4 & 4.77 $\pm$ 0.02 & 0.029 & 19.33 $\pm$ 0.07 & $-26.80$ & $0.06$ $\pm$ 0.41 \\
SDSSp J024457.19$-$010809.9 & 3.96 $\pm$ 0.01 & 0.030 & 18.46 $\pm$ 0.02 & $-27.38$ &  $-1.21$ $\pm$ 0.11\\
SDSSp J025019.78$+$004650.3 & 4.76 $\pm$ 0.01 & 0.052 & 19.64 $\pm$ 0.08 & $-26.49$ &  $-0.59$ $\pm$ 0.57\\
SDSSp J030025.23$+$003224.2 & 4.19 $\pm$ 0.01 & 0.095 & 19.96 $\pm$ 0.04 & $-25.97$ &  $-0.99$ $\pm$ 0.24\\
SDSSp J030707.46$-$001601.4 & 3.70 $\pm$ 0.01 & 0.063 & 20.04 $\pm$ 0.03 & $-25.69$ &  $-0.71$ $\pm$ 0.21\\
SDSSp J035214.33$-$001941.1 & 4.18 $\pm$ 0.01 & 0.167 & 19.51 $\pm$ 0.03 & $-26.42$ &  $-0.16$ $\pm$ 0.23\\
SDSSp J040550.26$+$005931.2 & 4.05 $\pm$ 0.01 & 0.444 & 19.27 $\pm$ 0.04& $-26.61$ &  $-$0.48 $\pm$ 0.21 \\
SDSSp J232717.99$+$000546.1 & 3.66 $\pm$ 0.01 & 0.036 & 19.93 $\pm$ 0.04& $-25.78$ &  $-$0.15 $\pm$ 0.20 \\
SDSSp J235053.55$-$004810.3 & 3.85 $\pm$ 0.01 & 0.027 & 19.80 $\pm$ 0.03 & $-26.00$ &  $-0.89$ $\pm$ 0.20\\
\enddata
\tablenotetext{}{Absolute magnitudes assume $H_0 = 50\rm\, km\, s^{-1}\,Mpc^{-1}$  and $\Omega_m = 1$}
\end{deluxetable}

\newpage

\begin{scriptsize}
\begin{deluxetable}{ccccrrc}
\tablenum{4}
\tablecolumns{7}
\tablecaption{Color Selected Sample of 182 deg$^2$ in Fall Equatorial Stripe}
\tablehead{quasar       & redshift &  AB$_{1450}$ &$M_{1450}$ & $\alpha$ & EW(Ly$\alpha$ $+$ NV) & p}
\startdata
SDSS J001950.06-004040.9 & 4.32 & 19.62 $\pm$ 0.05 & $-26.36$ &  $-0.02$ $\pm$ 0.62 & 53.5 & 0.90\\ 
SDSS J003525.29+004002.8 & 4.75 & 19.96 $\pm$ 0.08 & $-26.16$ &  $-0.84$ $\pm$ 0.60 & 82.4 & 0.99\\ 
SDSS J005922.65+000301.4 & 4.16 & 19.30 $\pm$ 0.03 & $-26.62$ &  $-1.09$ $\pm$ 0.33 & 77.1 & 0.78\\ 
SDSS J010619.25+004823.4 & 4.43 & 18.69 $\pm$ 0.04 & $-27.33$ &  $-0.40$ $\pm$ 0.31 & 69.6 & 0.74\\ 
SDSS J010822.70+001147.9 & 3.71 & 19.62 $\pm$ 0.04 & $-26.12$ &  $-0.19$ $\pm$ 0.24 & 74.6 & 0.98\\ 
SDSS J012019.99+000735.5 & 4.08 & 19.96 $\pm$ 0.04 & $-25.93$ &  $-0.52$ $\pm$ 0.30 & 82.9 & 0.95\\ 
SDSS J012403.78+004432.7 & 3.81 & 18.07 $\pm$ 0.02 & $-27.71$ &  $-0.44$ $\pm$ 0.11 & 74.2 & 0.98\\ 
SDSS J012650.77+011611.8 & 3.66 & 19.53 $\pm$ 0.03 & $-26.19$ &  $-0.35$ $\pm$ 0.19 & 76.0 & 0.84\\ 
SDSS J012700.69-004559.1 & 4.06 & 18.28 $\pm$ 0.02 & $-27.60$ &  $-0.66$ $\pm$ 0.13 & 93.5 & 0.98\\ 
SDSS J013108.19+005248.2 & 4.19 & 20.17 $\pm$ 0.04 & $-25.76$ & $0.11$ $\pm$ 0.34 & 71.6 & 0.42\\ 
SDSS J015048.83+004126.2 & 3.67 & 18.39 $\pm$ 0.02 & $-27.33$ &  $-0.52$ $\pm$ 0.12 & 47.9 & 0.82\\ 
SDSS J015339.61-001104.9 & 4.20 & 18.90 $\pm$ 0.02 & $-27.03$ &  $-1.32$ $\pm$ 0.33 & 69.3 & 0.55\\ 
SDSS J020427.81-011239.6 & 3.91 & 19.80 $\pm$ 0.03 & $-26.02$ &  $-0.83$ $\pm$ 0.17 & 71.1 & 0.84\\ 
SDSS J020731.68+010348.9 & 3.85 & 20.10 $\pm$ 0.04 & $-25.70$ &  $-1.00$ $\pm$ 0.21 & 60.2 & 0.56\\ 
SDSS J021043.17-001818.4 & 4.77 & 19.33 $\pm$ 0.07 & $-26.80$ & $0.06$ $\pm$ 0.41 & 42.5 & 0.97\\ 
SDSS J021102.72-000910.3 & 4.90 & 19.93 $\pm$ 0.10 & $-26.24$ &  $-0.99$ $\pm$ 0.70 & 60.9 & 1.00\\ 
SDSS J023231.40-000010.7 & 3.81 & 19.82 $\pm$ 0.03 & $-25.96$ &  $-0.44$ $\pm$ 0.18 & 48.6 & 0.91\\ 
SDSS J023908.98-002121.5 & 3.74 & 19.60 $\pm$ 0.03 & $-26.15$ &  $-0.78$ $\pm$ 0.16 & 65.8 & 0.89\\ 
SDSS J024457.19-010809.9 & 3.96 & 18.46 $\pm$ 0.02 & $-27.38$ &  $-1.21$ $\pm$ 0.11 & 87.9 & 0.64\\ 
SDSS J025019.78+004650.3 & 4.76 & 19.64 $\pm$ 0.08 & $-26.49$ &  $-0.59$ $\pm$ 0.57 & 42.8 & 0.99\\ 
SDSS J025112.44-005208.2 & 3.78 & 19.49 $\pm$ 0.03 & $-26.28$ & $0.05$ $\pm$ 0.49 & 75.1 & 0.89\\ 
SDSS J030025.23+003224.2 & 4.19 & 19.96 $\pm$ 0.04 & $-25.97$ &  $-0.99$ $\pm$ 0.24 & 114.7 & 0.91\\ 
SDSS J030707.46-001601.4 & 3.70 & 20.04 $\pm$ 0.03 & $-25.69$ &  $-0.71$ $\pm$ 0.21 & 60.3 & 0.76\\ 
SDSS J031036.85+005521.7 & 3.77 & 19.25 $\pm$ 0.03 & $-26.51$ &  $-0.64$ $\pm$ 0.15 & 45.0 & 0.83\\ 
SDSS J031036.97-001457.0 & 4.63 & 20.03 $\pm$ 0.07 & $-26.05$ &  $-0.02$ $\pm$ 0.65 & 56.9 & 0.97\\ 
SDSS J032608.12-003340.2 & 4.16 & 19.14 $\pm$ 0.03 & $-26.78$ &  $-0.33$ $\pm$ 0.18 & 58.8 & 1.00\\ 
SDSS J033829.31+002156.3 & 5.00 & 19.73 $\pm$ 0.12 & $-26.47$ &  $-0.81$ $\pm$ 0.76 & 71.5 & 0.98\\ 
SDSS J033910.53-003009.2 & 3.74 & 19.93 $\pm$ 0.03 & $-25.82$ &  $-1.17$ $\pm$ 0.20 & 74.7 & 0.69\\ 
SDSS J035214.33-001941.1 & 4.18 & 19.51 $\pm$ 0.03 & $-26.42$ &  $-0.16$ $\pm$ 0.23 & 108.3 & 0.99\\ 
SDSS J225419.23-000155.0 & 3.68 & 19.41 $\pm$ 0.03 & $-26.31$ &  $-0.95$ $\pm$ 0.17 & 87.2 & 0.78\\ 
SDSS J225452.88+004822.7 & 3.69 & 20.24 $\pm$ 0.04 & $-25.49$ &  $-1.51$ $\pm$ 0.27 & 95.6 & 0.31\\ 
SDSS J225529.09-003433.4 & 4.08 & 20.26 $\pm$ 0.06 & $-25.63$ &  $-1.15$ $\pm$ 0.37 & 82.8 & 0.52\\ 
SDSS J225759.67+001645.7 & 3.75 & 19.06 $\pm$ 0.02 & $-26.69$ &  $-0.58$ $\pm$ 0.15 & 72.3 & 0.96\\ 
SDSS J230323.77+001615.2 & 3.68 & 20.24 $\pm$ 0.04 & $-25.48$ &  $-0.77$ $\pm$ 0.26 & 107.1 & 0.56\\ 
SDSS J230639.65+010855.2 & 3.64 & 19.14 $\pm$ 0.03 & $-26.57$ &  $-1.38$ $\pm$ 0.15 & 57.5 & 0.45\\ 
SDSS J230952.29-003138.9 & 3.95 & 19.50 $\pm$ 0.03 & $-26.34$ &  $-0.72$ $\pm$ 0.18 & 63.8 & 0.94\\ 
SDSS J232208.22-005235.2 & 3.84 & 20.19 $\pm$ 0.04 & $-25.60$ &  $-1.18$ $\pm$ 0.24 & 59.6 & 0.73\\ 
SDSS J235053.55-004810.3 & 3.85 & 19.80 $\pm$ 0.03 & $-26.00$ &  $-0.89$ $\pm$ 0.20 & 52.8 & 0.76\\ 
SDSS J235718.35+004350.4 & 4.34 & 19.87 $\pm$ 0.05 & $-26.11$ &  $-1.08$ $\pm$ 0.65 & 47.1 & 0.38\\ 
\enddata
\tablenotetext{}{Absolute magnitudes assume $H_0 = 50\rm\, km\, s^{-1}\,Mpc^{-1}$  and $\Omega_m = 1$}
\end{deluxetable}

\end{scriptsize}

\newpage

\begin{deluxetable}{cccccc}
\tablenum{A1}
\tablecolumns{6}
\tablewidth{0pc}
\tablecaption{New SDSS Filter Characteristics of the SDSS 2.5m Imaging Camera, for 1.2 Air Mass}
\tablehead
{
 & $u'$ & $g'$ & $r'$ & $i'$ & $z'$ 
}
\startdata
$\lambda_{eff}$ (\AA) & 3561 & 4676 & 6176 & 7494 & 8873 \\
FWHM (\AA) & 560 & 1270 & 1130 & 1270 & 950
\enddata
\end{deluxetable}

\begin{deluxetable}{ccccc}
\tablenum{B1}
\tablecolumns{5}
\tablewidth{0pc}
\tablecaption{Summary of SDSS Photometric Runs}
\tablehead
{
Run & Date  & Stripe  & RA Range  & Seeing (arcsec)
}
\startdata
94  & 1998 Sep 19 & North & $335.38-58.92$ & 1.5   \\
125 & 1998 Sep 22 & South & $349.51 - 76.00$  & 1.6 \\
259 & 1998 Nov 17 & North & $7.27 - 96.04$  & $2.0 - 1.0$\\
273 & 1998 Nov 19 & South & $10.55 - 91.57$ & $1.6 - 1.4$ 
\enddata
\end{deluxetable}

\begin{scriptsize}
\begin{deluxetable}{cccccccr}
\tablenum{B2}
\tablecolumns{8}
\tablewidth{0pc}
\tablecaption{Positions and Photometry of Faint SDSS High-redshift Quasars}
\tablehead
{
SDSS name & redshift & $u^*$ & $g^*$ & $r^*$ & $i^*$ & $z^*$ & run
}
\startdata
SDSSp J004154.38$-$002955.9 & 3.82 $\pm$ 0.01 & 23.88 $\pm$ 0.70 & 21.66 $\pm$ 0.06 & 20.45 $\pm$ 0.03 & 20.35 $\pm$ 0.04 & 20.29 $\pm$ 0.16 &  94 \\ 
&  & 23.16 $\pm$ 1.00 & 21.91 $\pm$ 1.08 & 20.51 $\pm$  0.06 & 20.42 $\pm$  0.10 & 19.99 $\pm$  0.25 & 259 \\
SDSSp J005129.39$-$003644.7 & 3.71 $\pm$ 0.01 & 23.54 $\pm$ 0.60 & 21.75 $\pm$ 0.07 & 20.38 $\pm$ 0.03 & 20.23 $\pm$ 0.04 & 20.48 $\pm$ 0.19 &  94 \\ 
&  & 23.34 $\pm$ 1.06 & 21.76 $\pm$ 0.16 & 20.46 $\pm$  0.08 & 20.50 $\pm$  0.14 & 21.14 $\pm$  0.65& 259 \\
SDSSp J005348.66$-$002157.2 & 3.98 $\pm$ 0.02 & 24.42 $\pm$ 0.67 & 22.09 $\pm$ 0.12 & 20.43 $\pm$ 0.03 & 20.36 $\pm$ 0.05 & 20.60 $\pm$ 0.22 &  125 \\ 
&  & 23.26 $\pm$ 0.64 & 22.00 $\pm$ 0.14 & 20.65 $\pm$  0.05 & 20.73 $\pm$  0.08 & 20.90 $\pm$  0.41 & 273 \\
SDSSp J005452.86$-$001344.6 & 3.74 $\pm$ 0.01 & 23.58 $\pm$ 0.80 & 21.26 $\pm$ 0.06 & 20.18 $\pm$ 0.03 & 20.16 $\pm$ 0.04 & 20.13 $\pm$ 0.14 &  125 \\ 
&  & 23.96 $\pm$ 0.81 & 21.30 $\pm$ 0.07 & 20.30 $\pm$  0.05 & 20.21 $\pm$  0.06 & 20.22 $\pm$  0.26 & 273 \\
SDSSp J010905.81$+$001617.1 & 3.68 $\pm$ 0.01 & 24.16 $\pm$ 0.55 & 21.99 $\pm$ 0.07 & 20.79 $\pm$ 0.04 & 20.69 $\pm$ 0.06 & 21.34 $\pm$ 0.43 &  94 \\ 
&  & 24.38 $\pm$ 1.09 & 22.28 $\pm$ 0.18 & 20.91 $\pm$  0.09 & 20.94 $\pm$  0.16 & 20.74 $\pm$  0.40 & 259 \\
SDSSp J015015.58$+$004555.7 & 3.91 $\pm$ 0.01 & 24.47 $\pm$ 0.54 & 22.33 $\pm$ 0.10 & 20.80 $\pm$ 0.04 & 20.54 $\pm$ 0.05 & 20.51 $\pm$ 0.20 &  94 \\ 
&  & 23.40 $\pm$ 0.71 & 22.12 $\pm$ 0.13 & 20.78 $\pm$  0.06 & 20.58 $\pm$  0.07 & 20.36 $\pm$  0.16 & 259 \\
SDSSp J023749.33$+$005715.6 & 3.57 $\pm$ 0.01 & 22.85 $\pm$ 0.47 & 21.67 $\pm$ 0.07 & 20.60 $\pm$ 0.04 & 20.59 $\pm$ 0.06 & 20.57 $\pm$ 0.19 &  125 \\ 
&  & 23.20 $\pm$ 0.76 & 21.69 $\pm$ 0.10 & 20.50 $\pm$  0.05 & 20.46 $\pm$  0.08 & 20.55 $\pm$  0.34 & 273 \\
SDSSp J023935.25$+$010256.9 & 4.04 $\pm$ 0.01 & 23.46 $\pm$ 0.70 & 22.15 $\pm$ 0.10 & 20.86 $\pm$ 0.05 & 20.76 $\pm$ 0.06 & 20.29 $\pm$ 0.18 &  125 \\ 
&  & 24.16 $\pm$ 0.78 & 22.27 $\pm$ 0.17 & 20.67 $\pm$  0.06 & 20.52 $\pm$  0.09 & 21.03 $\pm$  0.50 & 273 \\
SDSSp J024347.37$-$010611.7 & 3.90 $\pm$ 0.01 & 22.69 $\pm$ 0.35 & 21.99 $\pm$ 0.11 & 20.39 $\pm$ 0.03 & 20.27 $\pm$ 0.04 & 20.04 $\pm$ 0.12 &  125 \\ 
&  & 24.54 $\pm$ 0.81 & 21.83 $\pm$ 0.10 & 20.27 $\pm$  0.05 & 20.27 $\pm$  0.06 & 19.94 $\pm$  0.18 & 273 \\
SDSSp J024434.87$+$000124.9 & 3.76 $\pm$ 0.01 & 23.19 $\pm$ 0.60 & 21.85 $\pm$ 0.08 & 20.49 $\pm$ 0.03 & 20.43 $\pm$ 0.04 & 20.36 $\pm$ 0.20 &  125 \\ 
&  & 23.64 $\pm$ 0.72 & 21.88 $\pm$ 0.11 & 20.63 $\pm$  0.05 & 20.43 $\pm$  0.06 & 20.25 $\pm$  0.25 & 273 \\
SDSSp J024452.33$-$003318.0 & 3.97 $\pm$ 0.01 & 23.44 $\pm$ 0.56 & 21.94 $\pm$ 0.07 & 20.40 $\pm$ 0.03 & 20.29 $\pm$ 0.04 & 20.06 $\pm$ 0.13 &  94 \\ 
&  & 23.79 $\pm$ 0.60 & 21.98 $\pm$ 0.08 & 20.38 $\pm$  0.05 & 20.28 $\pm$  0.05 & 20.10 $\pm$  0.15 & 259 \\
SDSSp J031427.92$+$002339.4 & 3.68 $\pm$ 0.01 & 24.25 $\pm$ 0.59 & 21.67 $\pm$ 0.06 & 20.44 $\pm$ 0.03 & 20.37 $\pm$ 0.04 & 20.22 $\pm$ 0.20 &  94 \\ 
&  & 24.68 $\pm$ 0.60 & 21.78 $\pm$ 0.08 & 20.38 $\pm$  0.05 & 20.45 $\pm$  1.08 & 20.32 $\pm$  0.20 & 259 \\
SDSSp J032459.10$-$005705.1 & 4.80 $\pm$ 0.02 & 23.12 $\pm$ 0.45 & 25.09 $\pm$ 0.72 & 22.73 $\pm$ 0.22 & 20.70 $\pm$ 0.06 & 20.31 $\pm$ 0.18 &  94 \\ 
&  & 23.99 $\pm$ 0.81 & 25.27 $\pm$ 0.84 & 22.51 $\pm$  0.27 & 20.51 $\pm$ 0.07 & 20.48 $\pm$  0.26 & 259 \\
SDSSp J033414.10$+$004056.6 & 4.33 $\pm$ 0.01 & 24.44 $\pm$ 0.62 & 23.31 $\pm$ 0.25 & 21.20 $\pm$ 0.06 & 20.82 $\pm$ 0.07 & 20.75 $\pm$ 0.26 &  94 \\ 
&  & 24.19 $\pm$ 0.82 & 23.44 $\pm$ 0.44 & 21.14 $\pm$  0.08 & 20.91 $\pm$  0.09 & 20.51 $\pm$  0.19 & 259 \\
SDSSp J033505.43$+$010337.2 & 3.58 $\pm$ 0.01 & 22.66 $\pm$ 0.37 & 22.35 $\pm$ 1.06 & 20.71 $\pm$ 0.04 & 20.55 $\pm$ 0.06 & 20.06 $\pm$ 0.15 &  94 \\ 
&  & 23.46 $\pm$ 0.83 & 22.23 $\pm$ 0.16 & 20.67 $\pm$  0.06 & 20.49 $\pm$  0.07 & 20.49 $\pm$  0.23 & 259 \\
SDSSp J042244.38$-$001247.8 & 4.09 $\pm$ 0.01 & 23.66 $\pm$ 0.69 & 22.40 $\pm$ 0.13 & 20.85 $\pm$ 0.04 & 20.46 $\pm$ 0.04 & 20.52 $\pm$ 0.17 &  125 \\ 
&  & 23.19 $\pm$ 0.56 & 22.46 $\pm$ 0.17 & 20.77 $\pm$  0.06 & 20.51 $\pm$  0.07 & 20.33 $\pm$  0.24 & 273 \\
SDSSp J042911.48$+$003501.7 & 3.74 $\pm$ 0.01 & 23.88 $\pm$ 0.66 & 23.26 $\pm$ 0.25 & 21.07 $\pm$ 0.05 & 20.65 $\pm$ 0.05 & 20.57 $\pm$ 0.17 &  125 \\ 
&  & 24.45 $\pm$ 0.70 & 23.33 $\pm$ 0.37 & 21.12 $\pm$  0.08 & 20.68 $\pm$  0.07 & 20.82 $\pm$  0.27 & 273 \\
SDSSp J043649.87$-$010612.9 & 3.99 $\pm$ 0.02 & 24.60 $\pm$ 0.55 & 21.84 $\pm$ 0.08 & 20.46 $\pm$ 0.03 & 20.24 $\pm$ 0.04 & 20.25 $\pm$ 0.13 &  125 \\ 
&  & 24.02 $\pm$ 0.05 & 21.76 $\pm$  0.09 & 20.39 $\pm$  0.05 & 20.12 $\pm$  0.05 & 20.32 $\pm$  0.24  & 273 
\enddata
\tablenotetext{}{Positions are in J2000.0 coordinates;
asinh magnitudes (Lupton, Gunn \& Szalay 1999) are
quoted; errors are statistical only.  For reference, zero flux
corresponds to asinh magnitudes of 23.40, 24.22, 23.98, 23.51, and
21.83 in $u^*, g^*, r^*, i^*$, and $z^*$, respectively.}
\end{deluxetable}
\end{scriptsize}

\newpage

\begin{scriptsize}
\begin{deluxetable}{cccccc}
\tablenum{B3}
\tablecolumns{6}
\tablewidth{0pc}
\tablecaption{Emission Line Properties of Faint SDSS High$-$redshift Quasars}
\tablehead
{
quasar           &      O$\,$VI 1034  & Ly$\alpha$ 1216 $+$ 1240 & O$\,$I+Si$\,$II 1306 & Si$\,$IV+O$\,$IV]  1402&
 C$\,$IV 1549 
}
\startdata
SDSSp J004154.38$-$002955.9 &  ... & 5890.3 $\pm$ 1.6 &  ... & 6756.3 $\pm$ 4.7 & 7467 $\pm$ 4.6\\ 
 &  ... &  60.6 $\pm$  2.2 &  ... &   7.1 $\pm$  1.1 &  29.4 $\pm$  5.9\\ 
SDSSp J005129.39$-$003644.7 &  ... & 5749.8 $\pm$ 0.5 & 6163.9 $\pm$ 2.4 & 6608.2 $\pm$ 3.1 & 7298 $\pm$ 1.4\\ 
 &  ... & 110.0 $\pm$  2.0 &   3.8 $\pm$  0.8 &  14.9 $\pm$  1.7 &  36.3 $\pm$  2.2\\ 
SDSSp J005348.66$-$002157.2 & 5147.4 $\pm$ 0.7 & 6064.3 $\pm$ 0.1 & 6509.9 $\pm$ 2.7 & 6959.9 $\pm$ 2.8 & 7746 $\pm$ 5.7\\ 
 &  16.8 $\pm$  0.5 &  59.2 $\pm$  0.7 &   4.5 $\pm$  0.9 &   5.5 $\pm$  0.8 &  50.4 $\pm$  4.6\\ 
SDSSp J005452.86$-$001344.6 & 4904.0 $\pm$ 1.3 & 5785.4 $\pm$ 0.2 & 6220.1 $\pm$ 5.0 & 6648.8 $\pm$ 2.2 & 7352 $\pm$ 0.9\\ 
 &  25.7 $\pm$  0.8 &  52.4 $\pm$  0.9 &   1.9 $\pm$  0.4 &   7.2 $\pm$  0.5 &  24.3 $\pm$  0.8\\ 
SDSSp J010905.81$+$001617.1 &  ... & 5791.6 $\pm$ 0.6 &  ... & 6547.1 $\pm$ 4.0 & 7251 $\pm$ 1.2\\ 
 &  ... & 110.6 $\pm$  0.7 &  ... &   8.1 $\pm$  0.5 &  13.0 $\pm$  0.6\\ 
SDSSp J015015.58$+$004555.7 & 5067.0 $\pm$ 0.6 &  ... &  ... & 6883.3 $\pm$ 2.2 & 7608 $\pm$ 1.6\\ 
 &   8.6 $\pm$  0.2 &  ... &  ... &  17.2 $\pm$  0.5 &  22.6 $\pm$  0.5\\ 
SDSSp J023749.33$+$005715.6 &  ... & 5563.4 $\pm$ 0.1 & 5977.1 $\pm$ 3.0 & 6389.9 $\pm$ 2.8 & 7082 $\pm$ 0.6\\ 
 &  ... &  21.3 $\pm$  0.2 &   2.5 $\pm$  0.4 &   7.0 $\pm$  0.6 &   8.3 $\pm$  0.3\\ 
SDSSp J023935.25$+$010256.9 &  ... & 6156.5 $\pm$ 0.2 & 6603.1 $\pm$ 2.0 & 7072.0 $\pm$ 2.2 & 7816 $\pm$ 1.1\\ 
 &  ... &  73.1 $\pm$  0.8 &   3.8 $\pm$  0.3 &  15.5 $\pm$  1.4 &  42.1 $\pm$  1.6\\ 
SDSSp J024347.37$-$010611.7 &  ... & 5979.2 $\pm$ 3.7 & 6415.2 $\pm$ 1.5 & 6864.2 $\pm$ 2.0 & 7553 $\pm$ 1.1\\ 
 &  ... & 108.8 $\pm$  1.7 &   3.9 $\pm$  0.3 &  12.1 $\pm$  0.7 &  11.7 $\pm$  0.4\\ 
SDSSp J024434.87$+$000124.9 &  ... & 5866.0 $\pm$ 6.4 &  ... &  ... & 7370 $\pm$ 16.6\\ 
 &  ... & 114.0 $\pm$ 16.1 &  ... &  ... &  31.3 $\pm$  5.8\\ 
SDSSp J024452.33$-$003318.0 &  ... & 6043.3 $\pm$ 1.9 &  ... &  ... &  ... \\ 
 &  ... &  55.9 $\pm$  5.7 &  ... &  ... &  ... \\ 
SDSSp J031427.92$+$002339.4 &  ... & 5697.2 $\pm$ 0.1 & 6117.6 $\pm$ 1.9 & 6548.0 $\pm$ 2.8 & 7248 $\pm$ 0.4\\
 &  ... &  53.2 $\pm$  0.8 &   2.1 $\pm$  0.2 &   5.7 $\pm$  0.4 &  24.3 $\pm$ 0.3\\
SDSSp J032459.10$-$005705.1 &  ... & 7042.9 $\pm$ 0.2 & 7578.3 $\pm$ 3.7 &  ... & ... \\
 &  ... &  42.8 $\pm$  0.3 &   3.7 $\pm$  0.7 &  ... &  ... \\
SDSSp J033414.10$+$004056.6 & 5512.8 $\pm$ 1.4 & 6493.1 $\pm$ 0.2 & 6956.1 $\pm$ 2.3 & 7458.7 $\pm$ 3.0 & 8258 $\pm$ 1.1\\
 &  22.3 $\pm$  1.0 &  56.8 $\pm$  1.1 &   3.1 $\pm$  0.3 &   8.4 $\pm$  0.8 & 26.3 $\pm$  1.2\\
SDSSp J033505.43$+$010337.2 &  ... &  ... & 5981.9 $\pm$ 1.2 & 6419.9 $\pm$ 2.9 & 7100 $\pm$ 1.7\\
 &  ... &  ... &   6.5 $\pm$  0.6 &   3.3 $\pm$  0.4 &  11.8 $\pm$  0.4\\
SDSSp J042244.38$-$001247.8 &  ... & 6186.5 $\pm$ 0.1 & 6662.3 $\pm$ 1.9 & 7124.7 $\pm$ 2.6 & 7875 $\pm$ 0.9\\ 
 &  ... &  49.7 $\pm$  0.8 &   2.2 $\pm$  0.2 &  10.0 $\pm$  0.6 &  23.9 $\pm$  0.5\\ 
SDSSp J042911.48$+$003501.7 & 4903.0 $\pm$ 0.9 & 5768.2 $\pm$ 0.1 & 6182.5 $\pm$ 1.3 & 6634.0 $\pm$ 3.4 & 7342 $\pm$ 0.3\\ 
 &  29.4 $\pm$  2.0 & 116.3 $\pm$  2.1 &   6.0 $\pm$  0.4 &   8.1 $\pm$  0.9 &  46.3 $\pm$  0.6\\ 
SDSSp J043649.87$-$010612.9 &  ... & 6151.5 $\pm$ 1.0 & 6561.5 $\pm$ 4.6 & 6996.2 $\pm$ 6.0 & 7722 $\pm$ 2.8\\ 
 &  ... &  81.0 $\pm$  1.0 &   5.2 $\pm$  1.7 &   8.6 $\pm$  0.8 &  25.7 $\pm$  1.1\\ 
\enddata
\tablenotetext{}{The two entries in each line are the central wavelength and rest frame equivalent width
from the Gaussian fit to the line profile, both measured in
\AA{}ngstroms.}
\end{deluxetable}
\end{scriptsize}
\clearpage
\newpage

\begin{deluxetable}{ccccc}
\tablenum{B4}
\tablecolumns{5}
\tablecaption{Continuum Properties of Faint SDSS High$-$redshift Quasars}
\tablehead{quasar       & redshift & $E(B-V)$ & AB$_{1450}$ &$M_{1450}$}
\startdata
SDSSp J004154.38$-$002955.9 & 3.82 $\pm$ 0.01 & 0.021 & 20.50 $\pm$ 0.04& $-$25.28  \\ 
SDSSp J005129.39$-$003644.7 & 3.71 $\pm$ 0.01 & 0.051 & 20.49 $\pm$ 0.04& $-$25.25  \\ 
SDSSp J005348.66$-$002157.2 & 3.98 $\pm$ 0.02 & 0.026 & 20.55 $\pm$ 0.05& $-$25.30  \\ 
SDSSp J005452.86$-$001344.6 & 3.74 $\pm$ 0.01 & 0.025 & 20.31 $\pm$ 0.04& $-$25.44  \\ 
SDSSp J010905.81$+$001617.1 & 3.68 $\pm$ 0.01 & 0.026 & 20.89 $\pm$ 0.05& $-$24.83  \\ 
SDSSp J015015.58$+$004555.7 & 3.91 $\pm$ 0.01 & 0.029 & 20.58 $\pm$ 0.05& $-$25.24  \\ 
SDSSp J023749.33$+$005715.6 & 3.57 $\pm$ 0.01 & 0.032 & 20.67 $\pm$ 0.05& $-$25.00  \\ 
SDSSp J023935.25$+$010256.9 & 4.04 $\pm$ 0.01 & 0.031 & 20.95 $\pm$ 0.06& $-$24.92  \\ 
SDSSp J024347.37$-$010611.7 & 3.90 $\pm$ 0.01 & 0.032 & 20.38 $\pm$ 0.04& $-$25.44  \\ 
SDSSp J024434.87$+$000124.9 & 3.76 $\pm$ 0.01 & 0.030 & 20.64 $\pm$ 0.04& $-$25.12  \\ 
SDSSp J024452.33$-$003318.0 & 3.97 $\pm$ 0.01 & 0.030 & 20.36 $\pm$ 0.04& $-$25.48  \\ 
SDSSp J031427.92$+$002339.4 & 3.68 $\pm$ 0.01 & 0.098 & 20.33 $\pm$ 0.04& $-$25.39  \\ 
SDSSp J032459.10$-$005705.1 & 4.80 $\pm$ 0.02 & 0.099 & 20.53 $\pm$ 0.14& $-$25.61  \\ 
SDSSp J033414.10$+$004056.6 & 4.33 $\pm$ 0.01 & 0.132 & 20.69 $\pm$ 0.10& $-$25.29  \\ 
SDSSp J033505.43$+$010337.2 & 3.58 $\pm$ 0.01 & 0.100 & 20.46 $\pm$ 0.05& $-$25.22  \\ 
SDSSp J042244.38$-$001247.8 & 4.09 $\pm$ 0.01 & 0.071 & 20.50 $\pm$ 0.05& $-$25.40  \\ 
SDSSp J042911.48$+$003501.7 & 3.74 $\pm$ 0.01 & 0.075 & 21.02 $\pm$ 0.06& $-$24.73  \\ 
SDSSp J043649.87$-$010612.9 & 3.99 $\pm$ 0.02 & 0.037 & 20.37 $\pm$ 0.04& $-$25.48  \\ 
\enddata
\tablenotetext{}{Absolute magnitudes assume $H_0 = 50\rm\, km\, s^{-1}\,Mpc^{-1}$  and $\Omega_m = 1$}
\end{deluxetable}

\begin{figure}
\vspace{-3cm}

\epsfysize=400pt \epsfbox{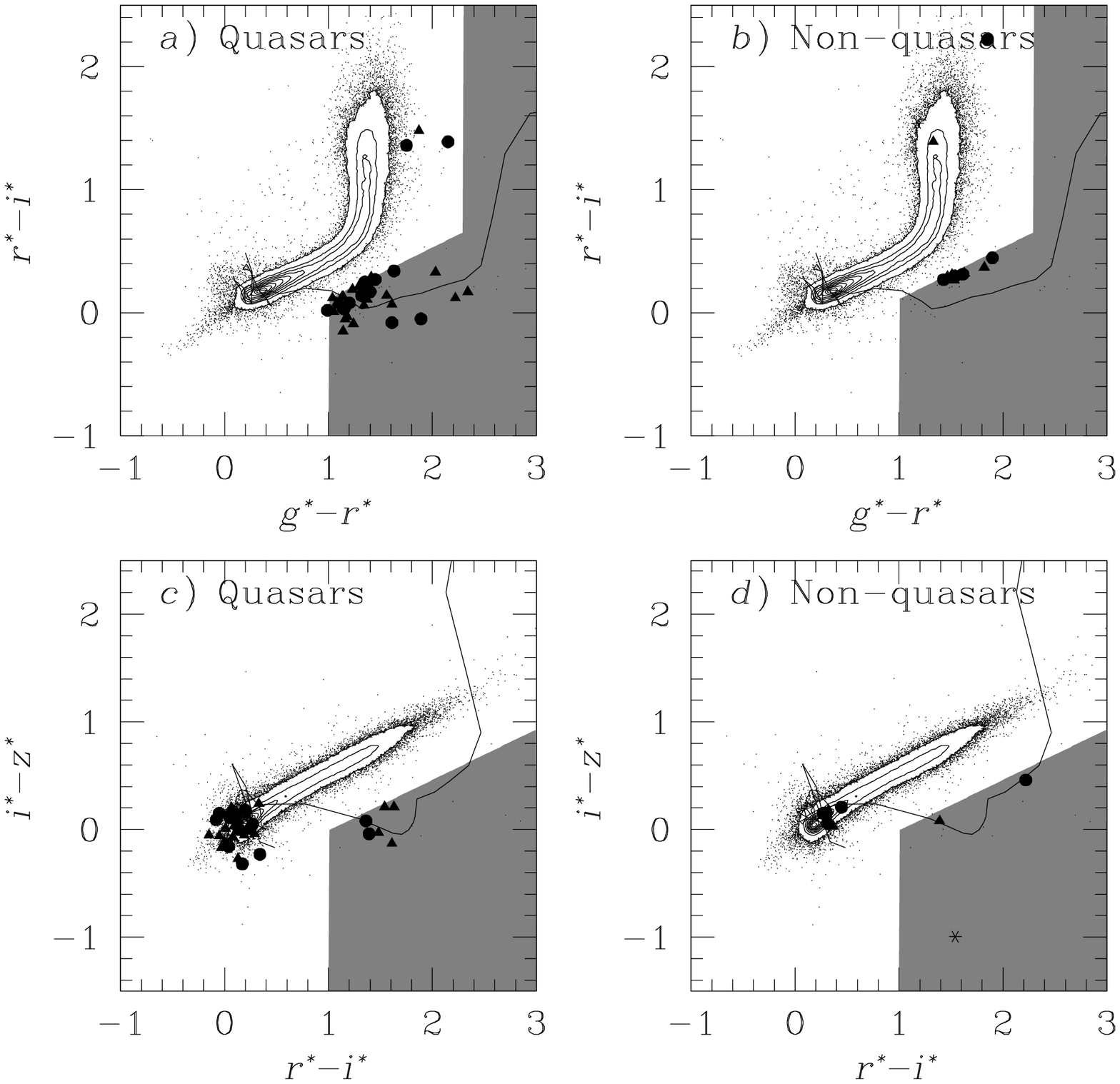}
\vspace{1cm}
Figure 1. Color-color diagrams for all stellar objects with $i^* < 20.2$ 
 in 25 square degrees of SDSS imaging data.
The inner parts of the diagrams are shown as contours,
linearly spaced in the density of stars in color-color space.
The shaded areas on the $g^*- r^*$ vs. $r^* - i^*$ and
the $r^* - i^*$ vs. $i^* - z^*$ diagrams represent the
selection criteria used to select quasar candidates.
The solid line is the median track of simulated quasar colors
as a function of redshift (adapted from Fan 1999).  
(a) and (c) show the colors of 39 quasars 
at $z>3.6$ in the complete sample. 
The circles are new quasars presented in this paper; the triangles
are quasars presented in Fan et al. (1999a) and Schneider et al. (2000b).
(b) and (d) shows the colors of 14 non-quasars that
satisfy the selection criteria. 
The circles are objects identified as galaxies; 
the triangles are stars; and the asterisks are objects with
unidentified spectra.
\end{figure}

\begin{figure}
\vspace{-0.5cm}

\epsfysize=500pt \epsfbox{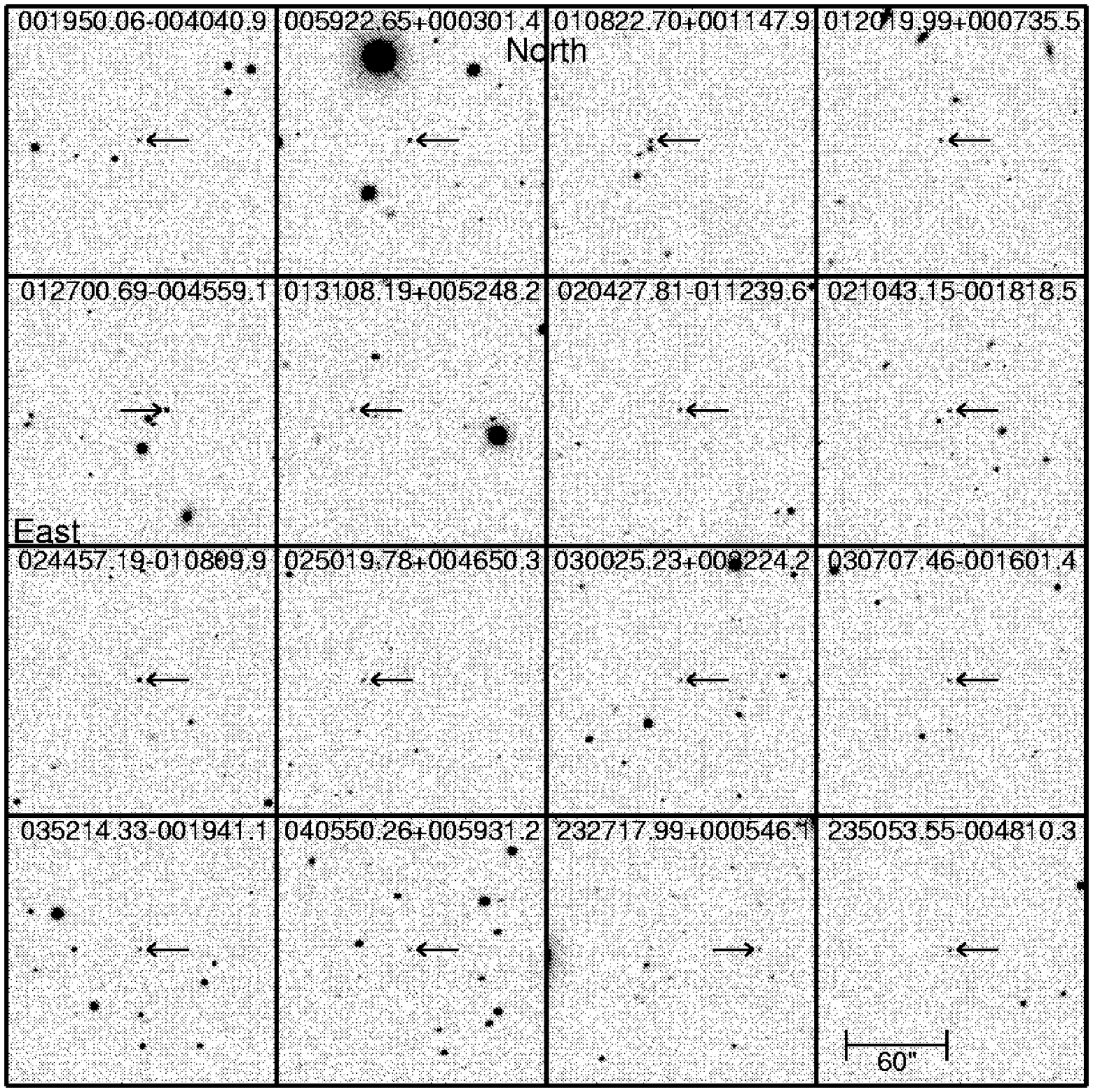}
\vspace{1cm}
Figure 2. Finding charts for the 16 bright SDSS quasars.
The data are $160'' \times 160''$ SDSS images in the $i'$ band
(54.1 sec exposure time). Most of them are re-constructed
from the atlas images and binned background from the SDSS database.
North is up; East is to the left.
\end{figure}

\clearpage
\begin{figure}
\vspace{-3cm}

\epsfysize=600pt \epsfbox{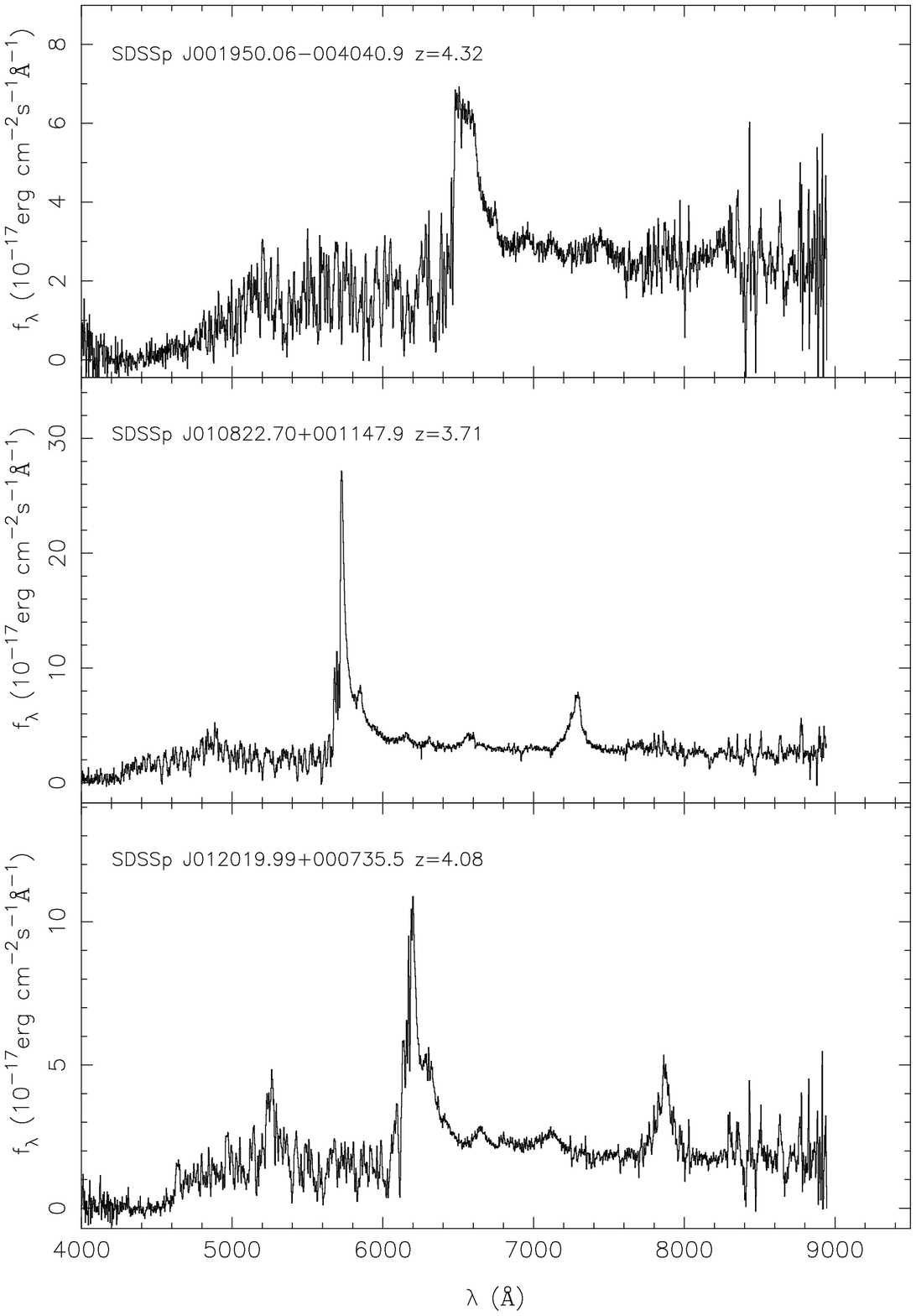}
\vspace{1cm}
Figure 3. Spectra of 15 bright SDSS quasars.
Most are Keck II/LRIS spectra with a dispersion of 2 \AA/pixel.
The spectrum of SDSSp J024457.19--010809.9 (PSS 0244--0108) was
taken with the ARC 3.5m at a dispersion of 6\AA/pixel.

\end{figure}

\begin{figure}
\vspace{-3cm}

\plotone{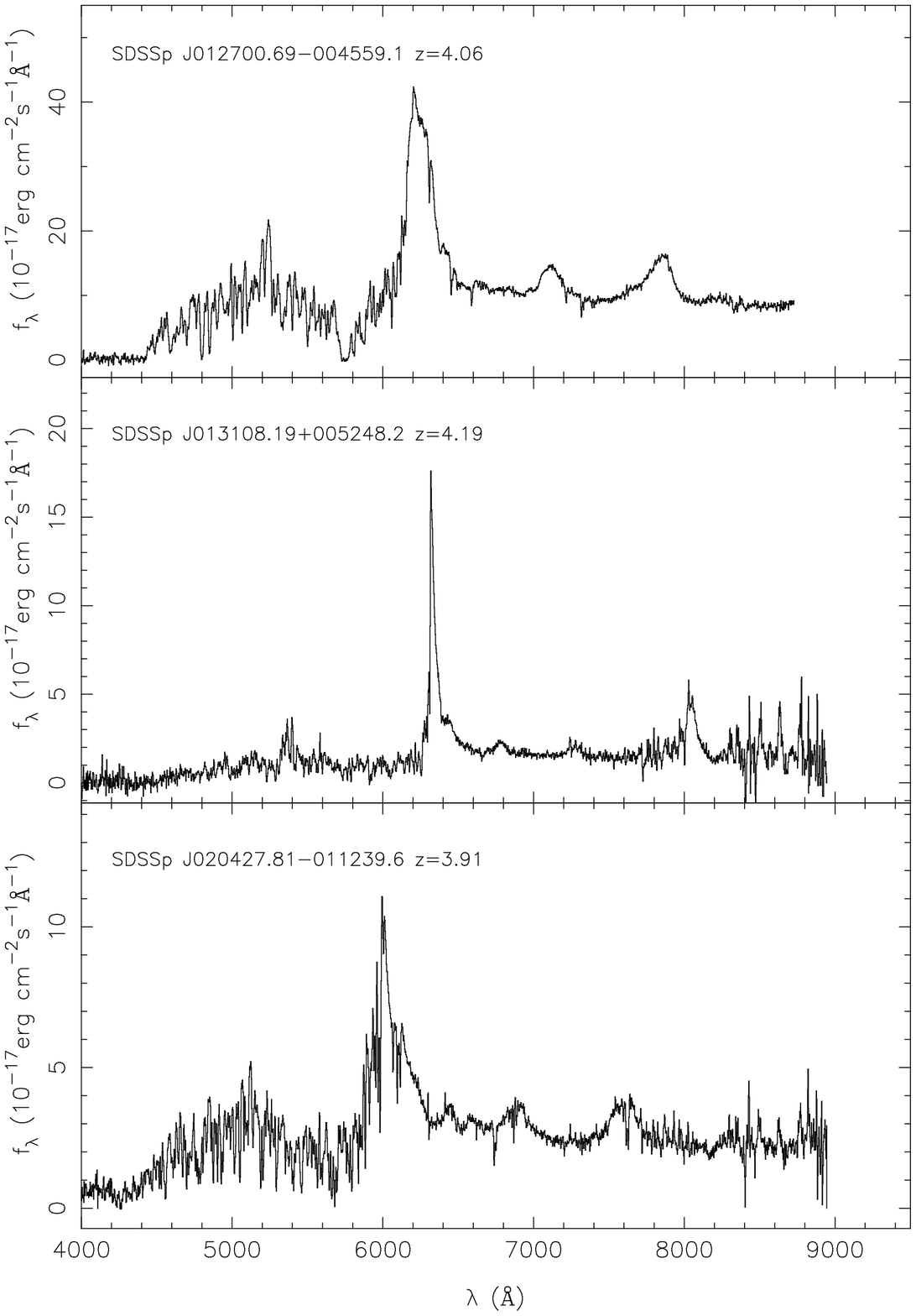}
\vspace{1cm}
Figure 3. Continued
\end{figure}
\newpage

\begin{figure}
\vspace{-3cm}

\plotone{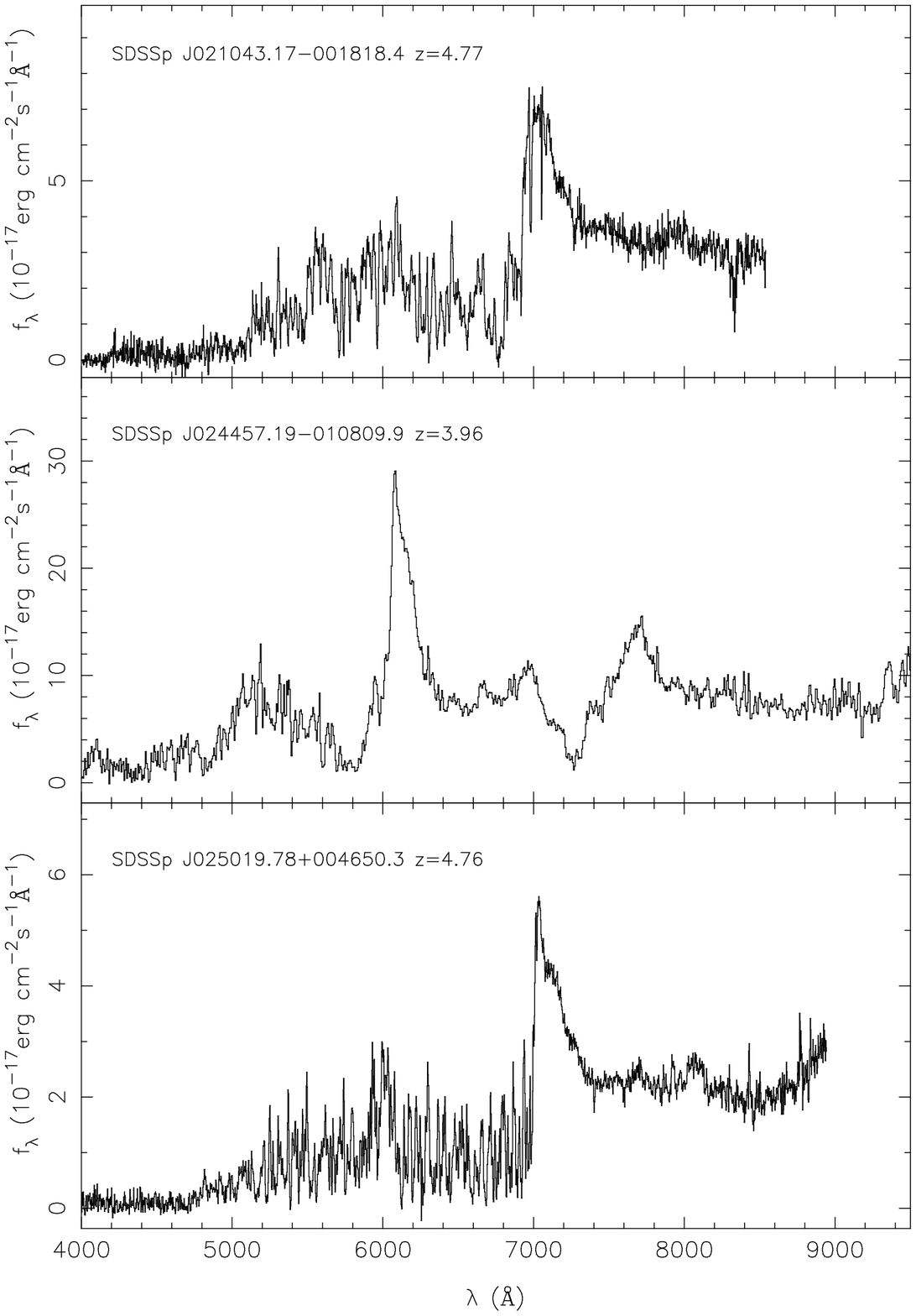}
\vspace{1cm}
Figure 3. Continued
\end{figure}
\newpage

\begin{figure}
\vspace{-3cm}

\plotone{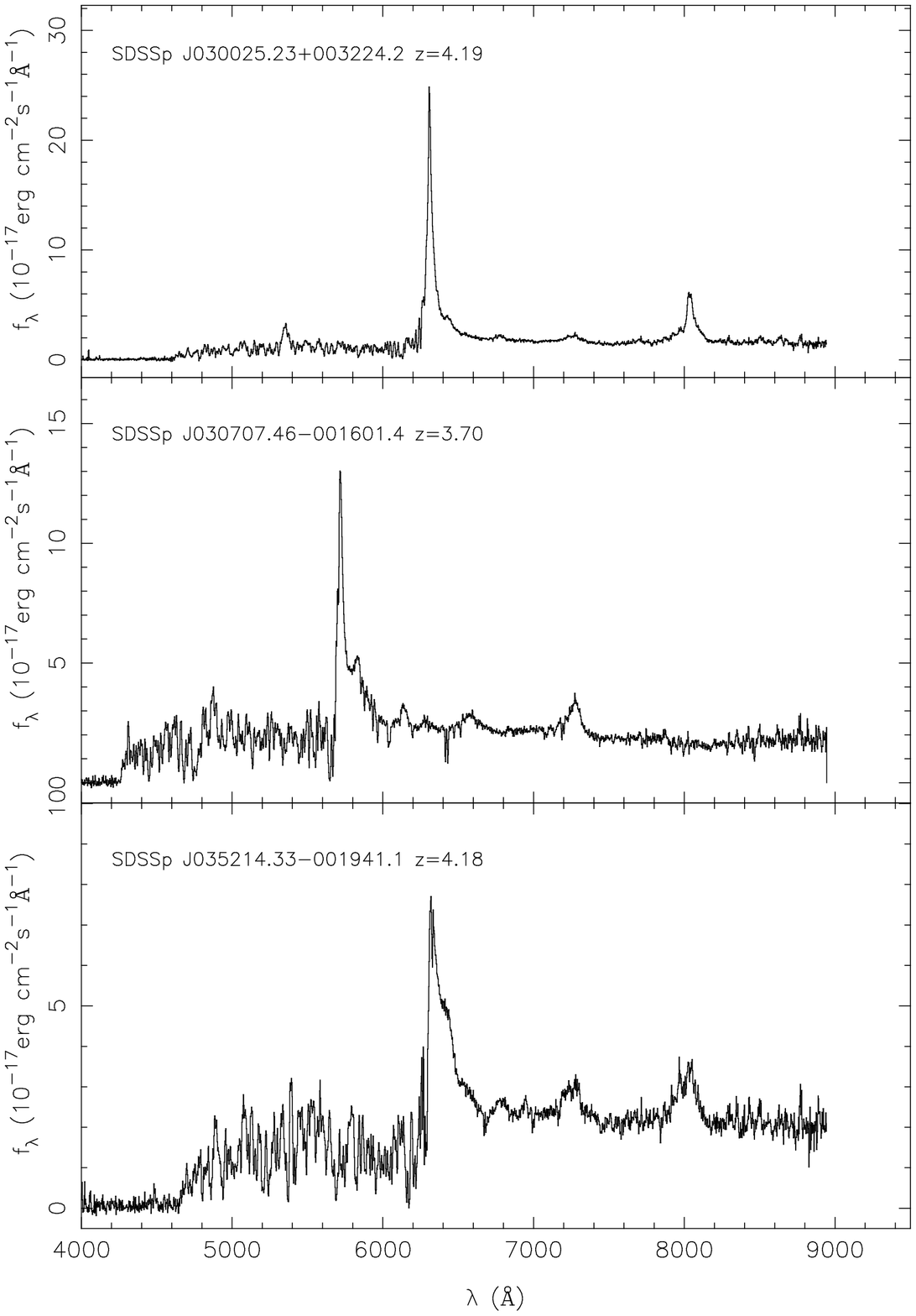}
\vspace{1cm}
Figure 3. Continued
\end{figure}
\newpage

\begin{figure}
\vspace{-3cm}

\plotone{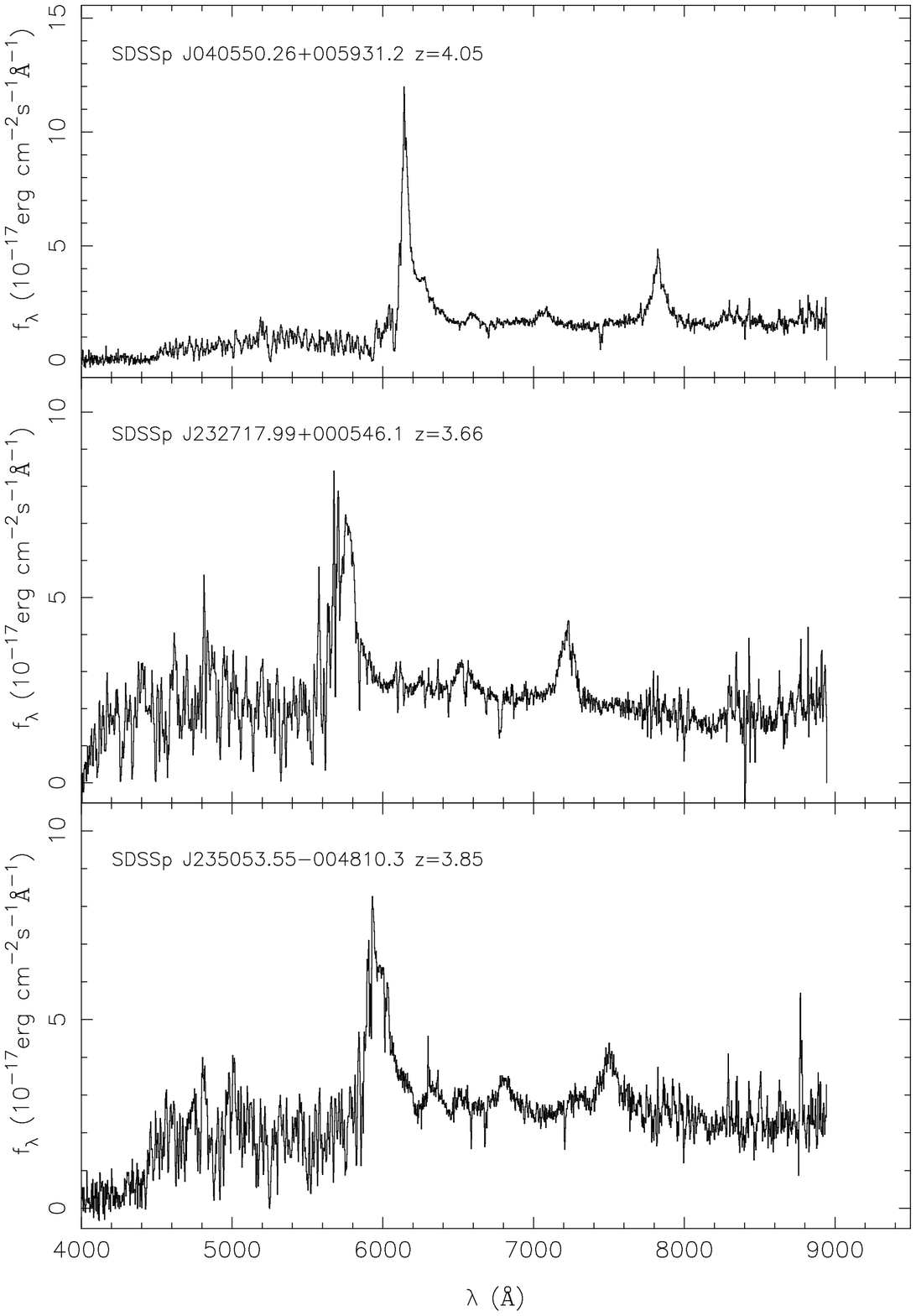}
\vspace{1cm}
Figure 3. Continued
\end{figure}
\newpage

\begin{figure}
\vspace{-3cm}

\epsfysize=600pt \epsfbox{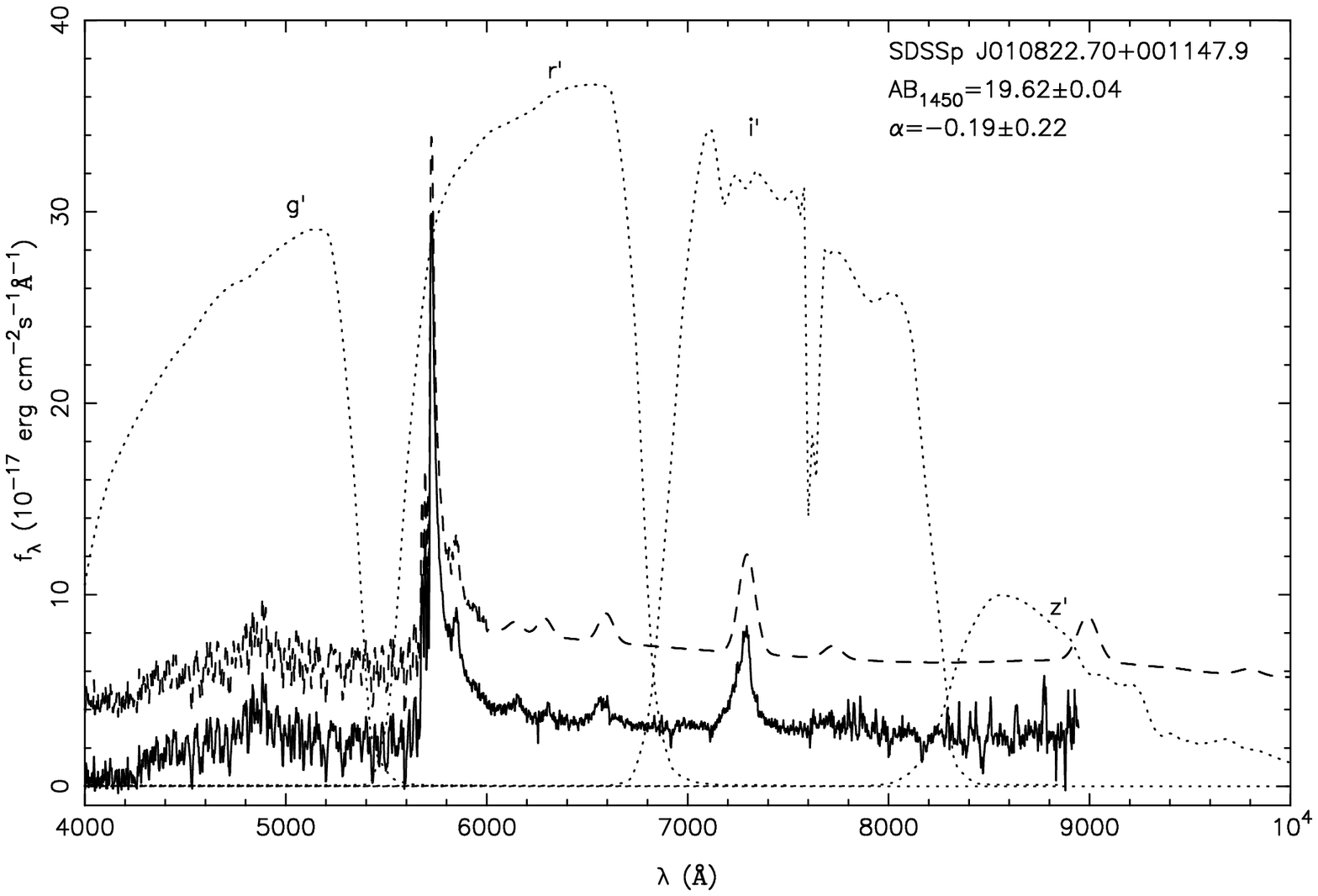}

\vspace{1cm}
Figure 4. An illustration of the determination of the continuum properties.
The solid line is the observed Keck spectrum, after applying a linear correction
function to reproduce the SDSS $r^*$ and $i^*$ magnitudes.
The dashed line is the spectral model (with an offset) that provides the
smallest $\chi^2$ for the SDSS $r^*$, $i^*$ and $z^*$ measurements.
The dotted lines are the system response functions for the SDSS passbands
(see Fig A-1).

\end{figure}
\newpage

\begin{figure}
\vspace{-5cm}

\epsfysize=600pt \epsfbox{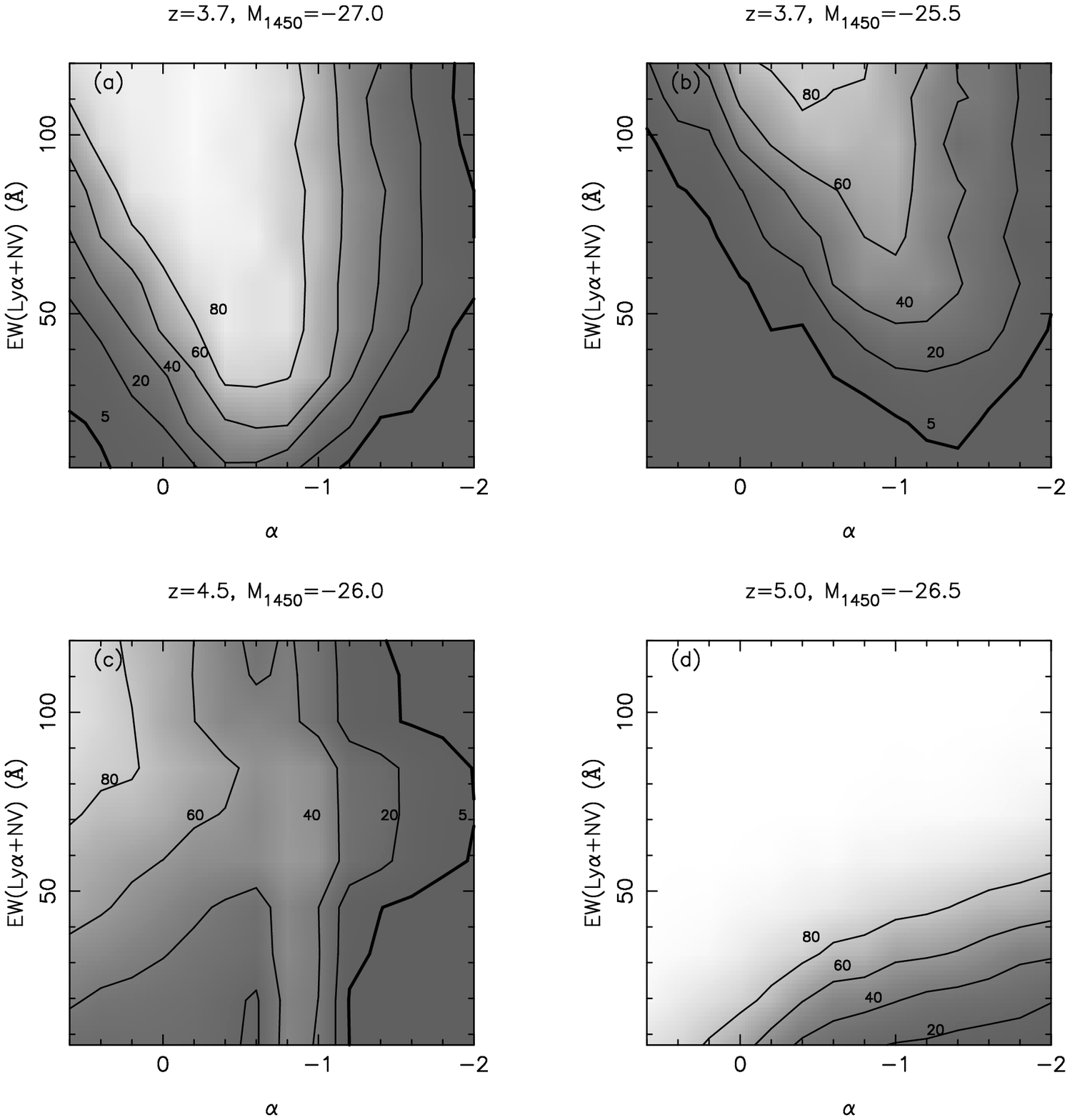}
\vspace{1cm}
Figure 5. Quasar selection probability as a function of
continuum slope and emission line strength at several different redshifts and
luminosities.
\end{figure}
\newpage

\begin{figure}
\vspace{-1.5cm}

\epsfysize=600pt \epsfbox{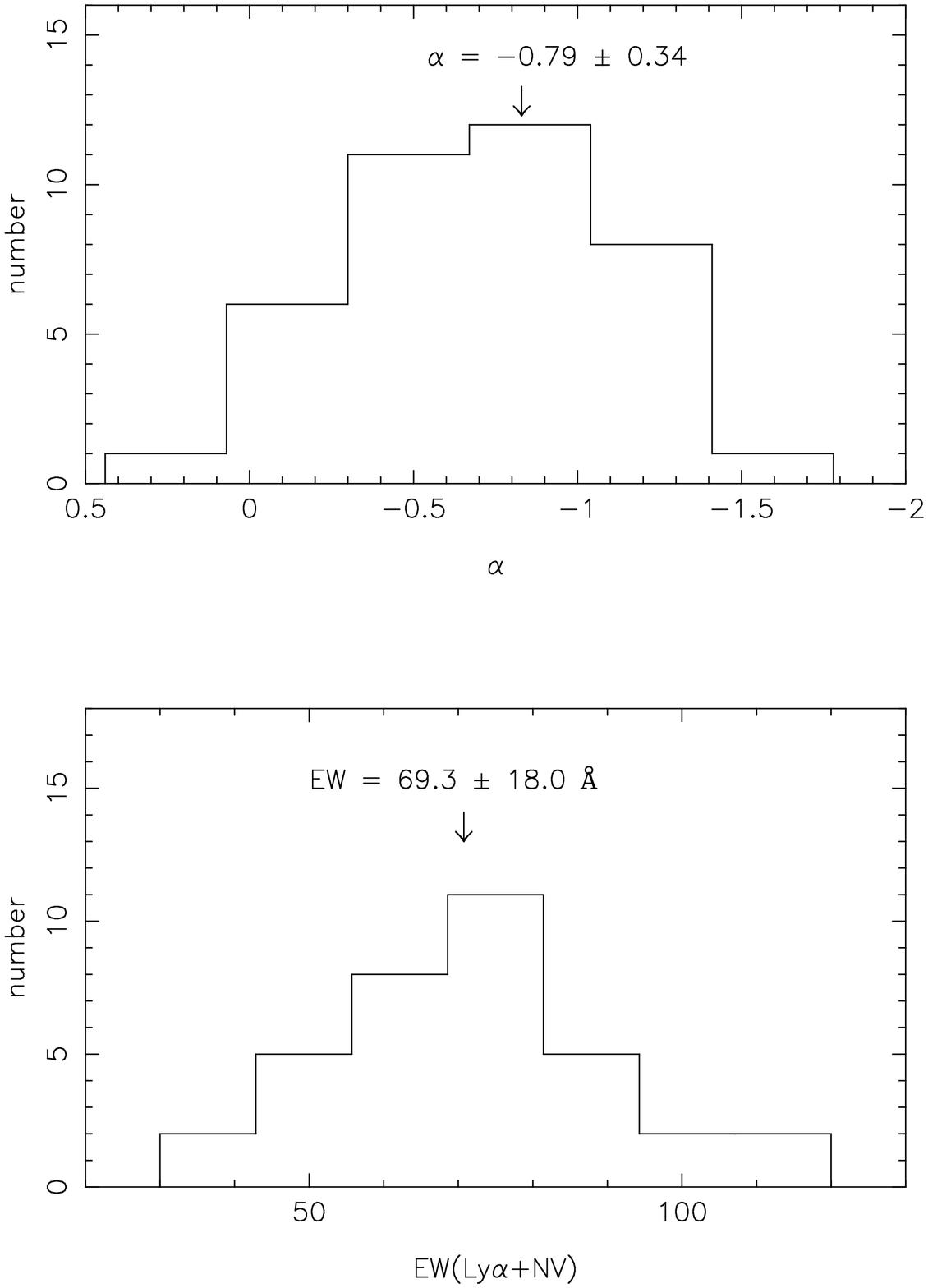}
Figure 6. Distribution of the quasar continuum power-law index $\alpha$ (upper
 panel)
and the rest-frame equivalent width of Ly$\alpha$+NV (lower panel) for the
Fall Equatorial Stripe sample.
The lower panel only includes the 35 non-BAL quasars.
Note that the average $\alpha$ from the maximum likelihood fit is not centered
on the observed distribution. This is due to the fact that the selection criteria
are less sensitive to objects with larger $\alpha$ (see Figure 5).

\end{figure}
\newpage

\begin{figure}
\vspace{-5cm}

\epsfysize=600pt \epsfbox{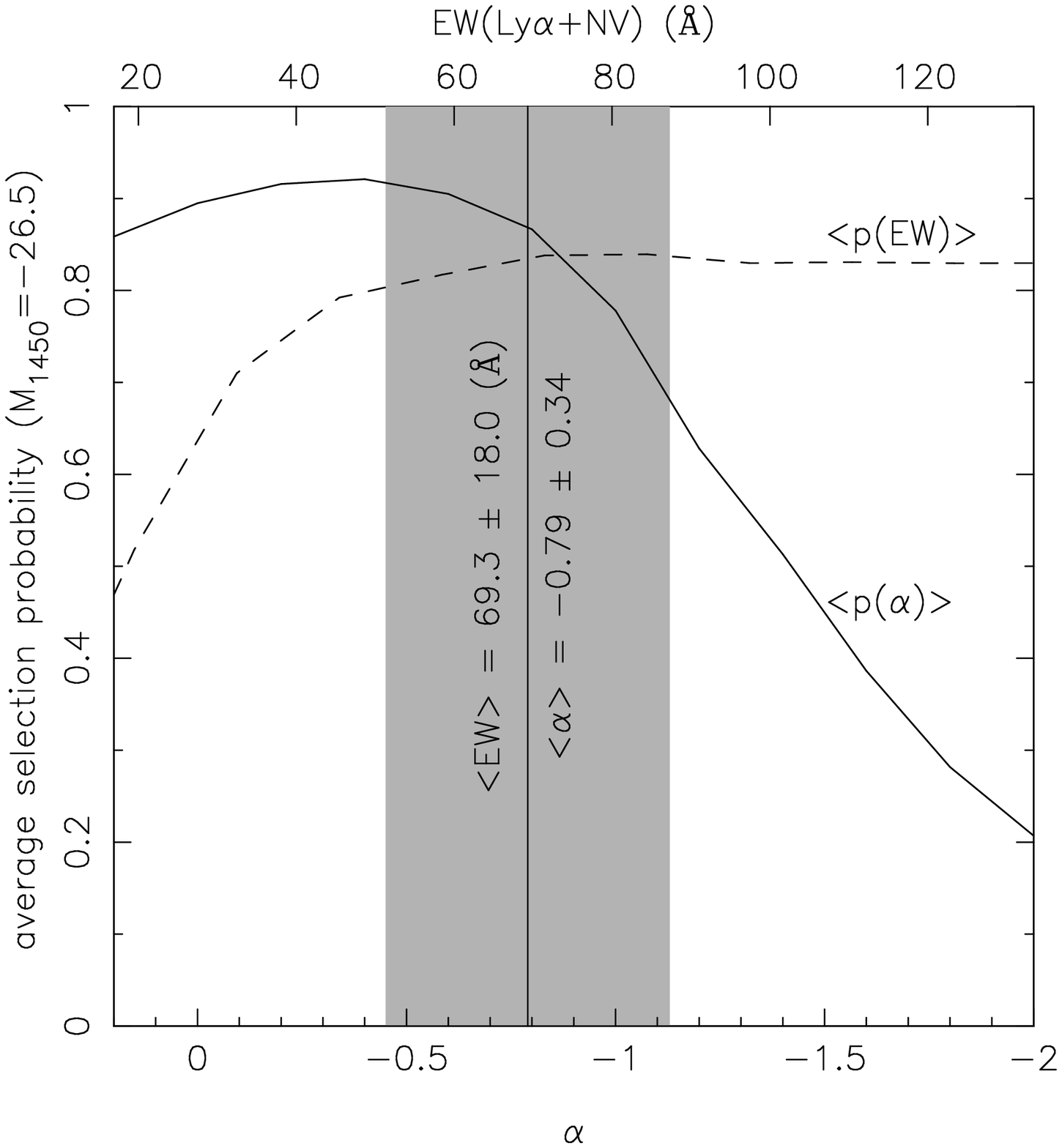}

\vspace{1cm}
Figure 7. The average selection probability
for $M_{1450} = -26.5$ and redshift range $3.6 < z < 5.0$.
The solid line is the average probability as a function of
continuum slope, and the dashed line is the average probability
as a function of the emission
line strength.
The shaded area shows the $1-\sigma$ scatter of the distributions
of continuum slope and emission line strength measured from the
SDSS sample.
\end{figure}

\begin{figure}
\vspace{-1.5cm}

\epsfysize=600pt \epsfbox{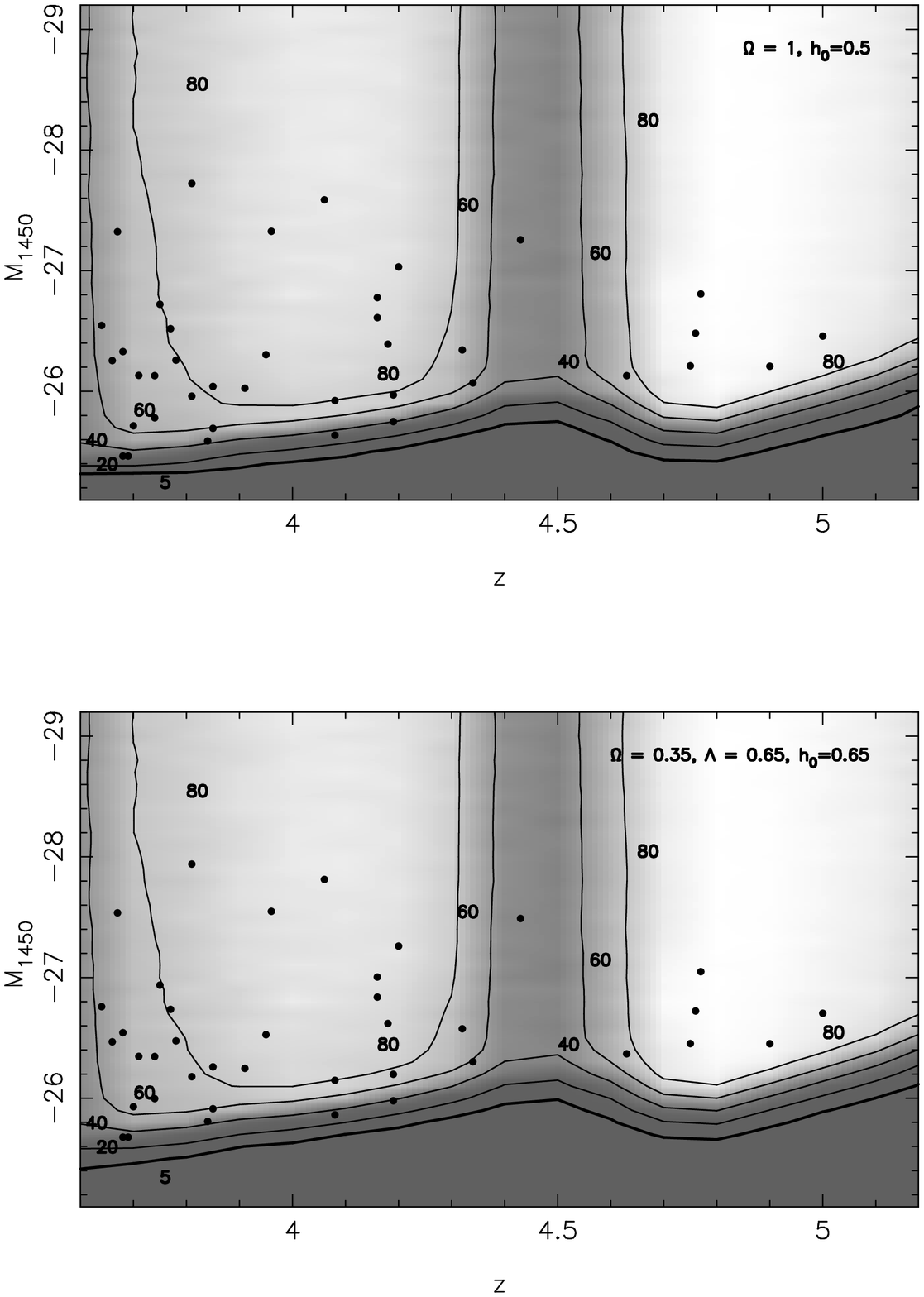}
\vspace{1cm}
Figure 8. The selection probability of high-redshift  quasars
in the Fall Equatorial Stripe sample as a function of redshift and
luminosity, for the $\Omega=1$ and the $\Lambda$ models. 
The probability contours of 5\%, 20\%, 40\%, 60\% and 80\% are shown.
The large dots represent the locations of the 39 quasars in the sample.

\end{figure}
\newpage

\begin{figure}

\vspace{-5cm}
\epsfysize=600pt \epsfbox{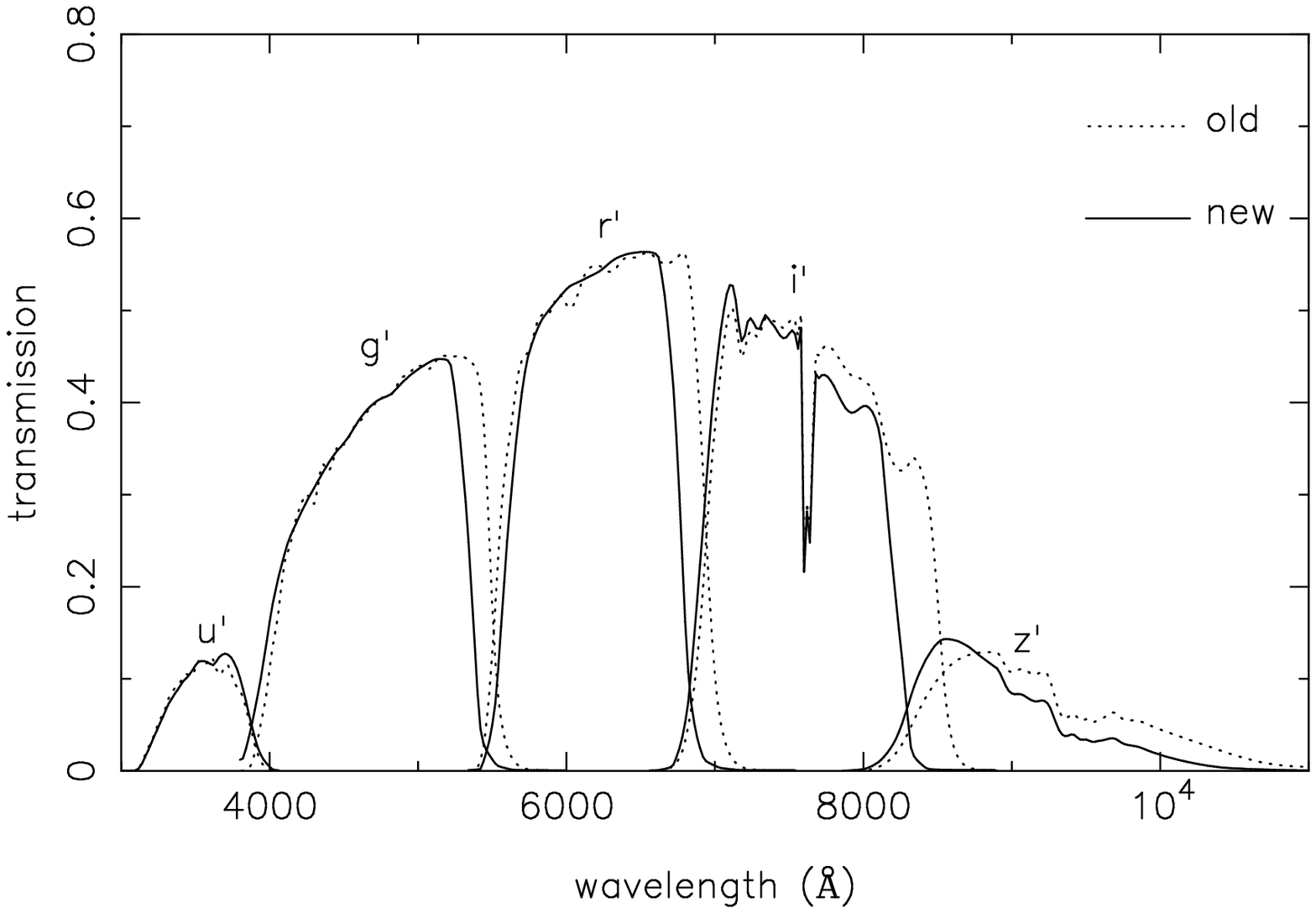}
\vspace{1cm}
Figure A1. The SDSS photometric system response function at 1.2 airmass.
The solid lines shows the new measurements in Spring 2000;
the dotted lines are the old measurements before the installation of
the SDSS camera in 1997.
Note that the red-edge cutoffs in the $g'$, $r'$, $i'$ and $z'$ filters have moved
to shorter wavelengths. 
 
\end{figure}

\begin{figure}
\epsfysize=500pt \epsfbox{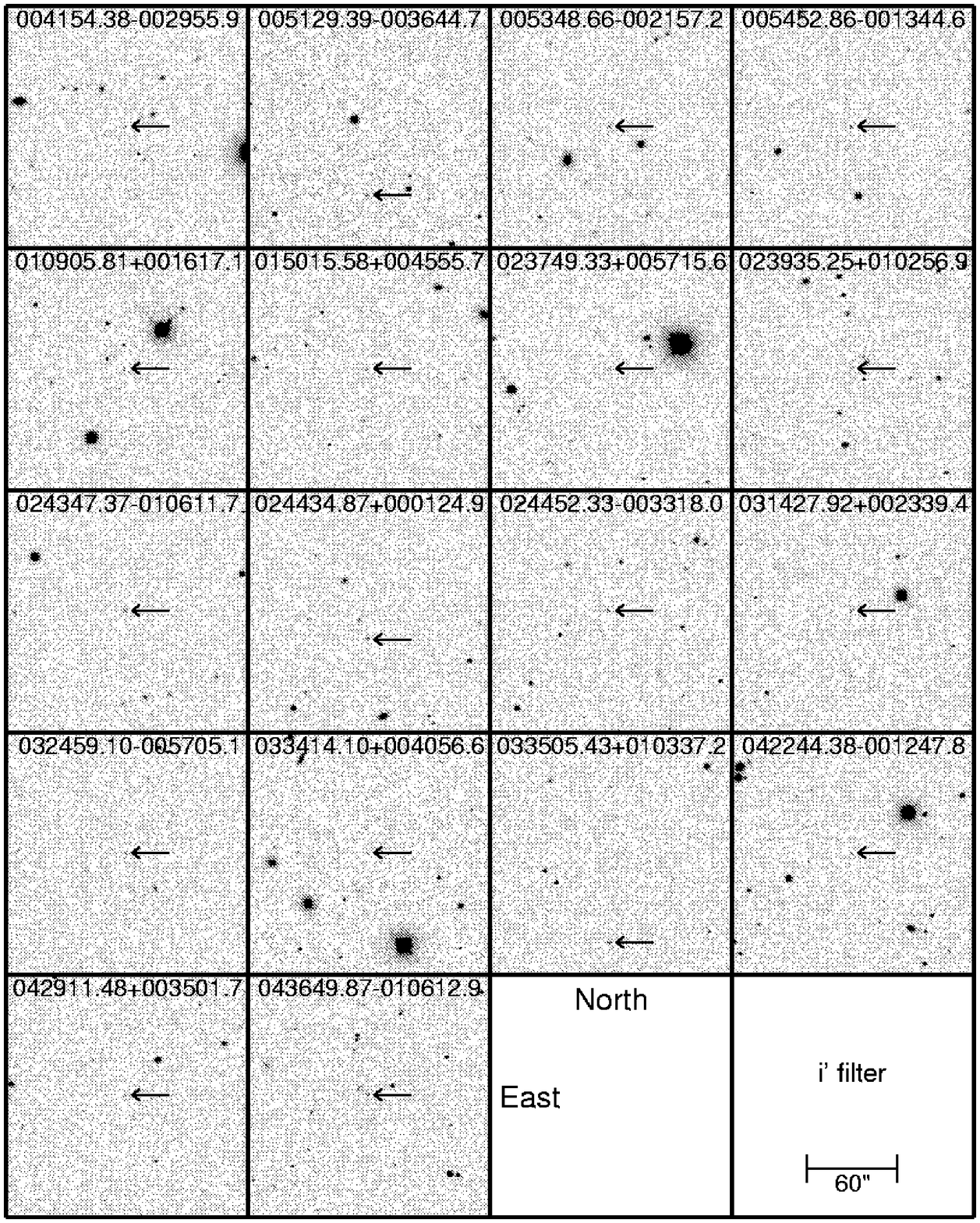}
\vspace{1cm}
Figure B1. Finding charts for the 18 faint SDSS quasars.
The data are $160'' \times 160''$ SDSS images in the $i'$ band
(54.1 sec exposure time).
North is up; East is to the left.
\end{figure}

\begin{figure}

\vspace{-3cm}

\epsfysize=600pt \epsfbox{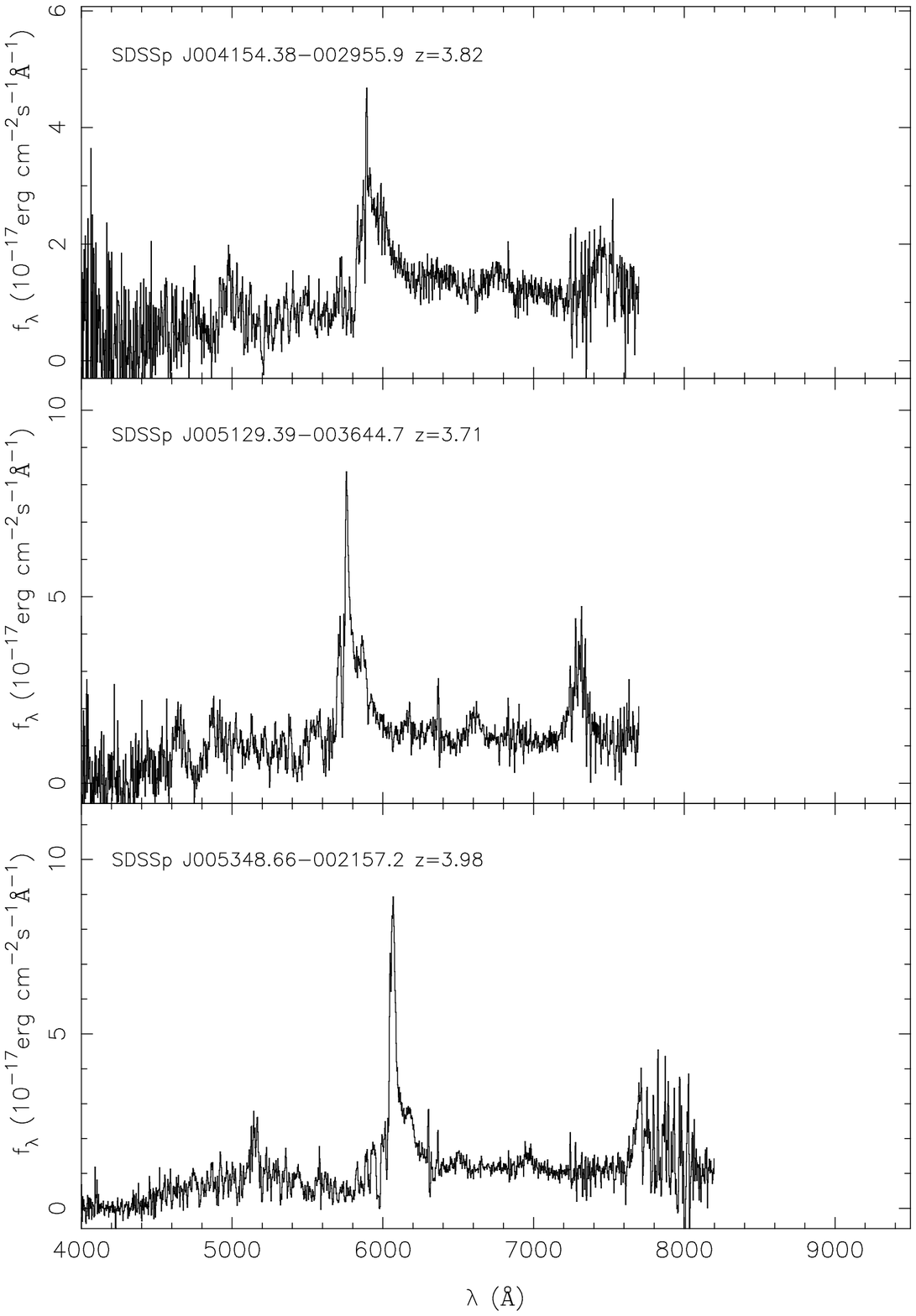}
\vspace{1cm}
Figure B2. The spectra of 18 faint SDSS high-redshift quasars obtained
with the KeckII and the ARC 3.5m telescopes (for more details, see captions of Figure 3).

\end{figure}
\newpage

\begin{figure}
\vspace{-3cm}

\plotone{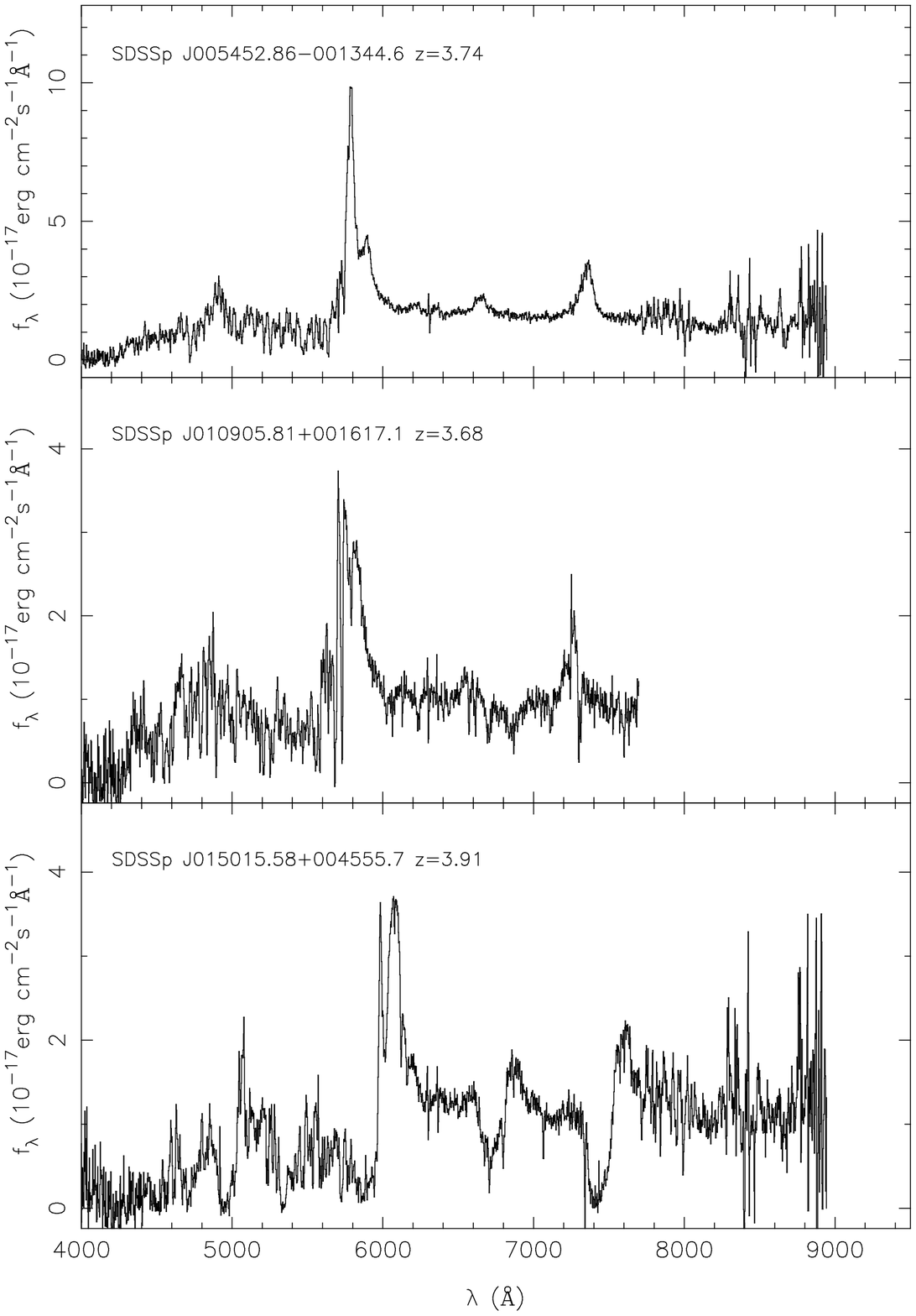}
\vspace{1cm}
Figure B2. Continued

\end{figure}
\newpage

\begin{figure}
\vspace{-3cm}

\plotone{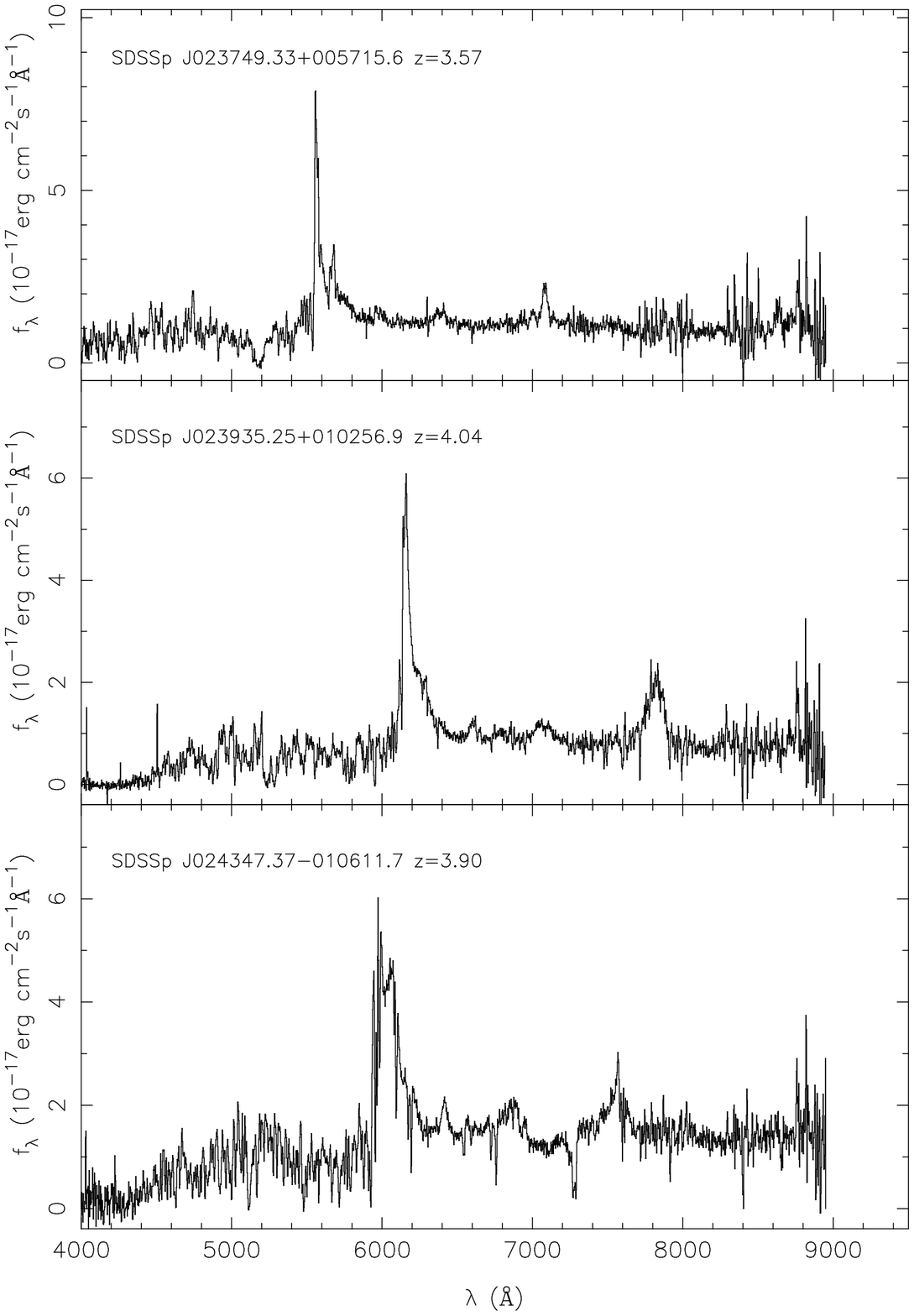}
\vspace{1cm}
Figure B2. Continued

\end{figure}
\newpage

\begin{figure}
\vspace{-3cm}

\plotone{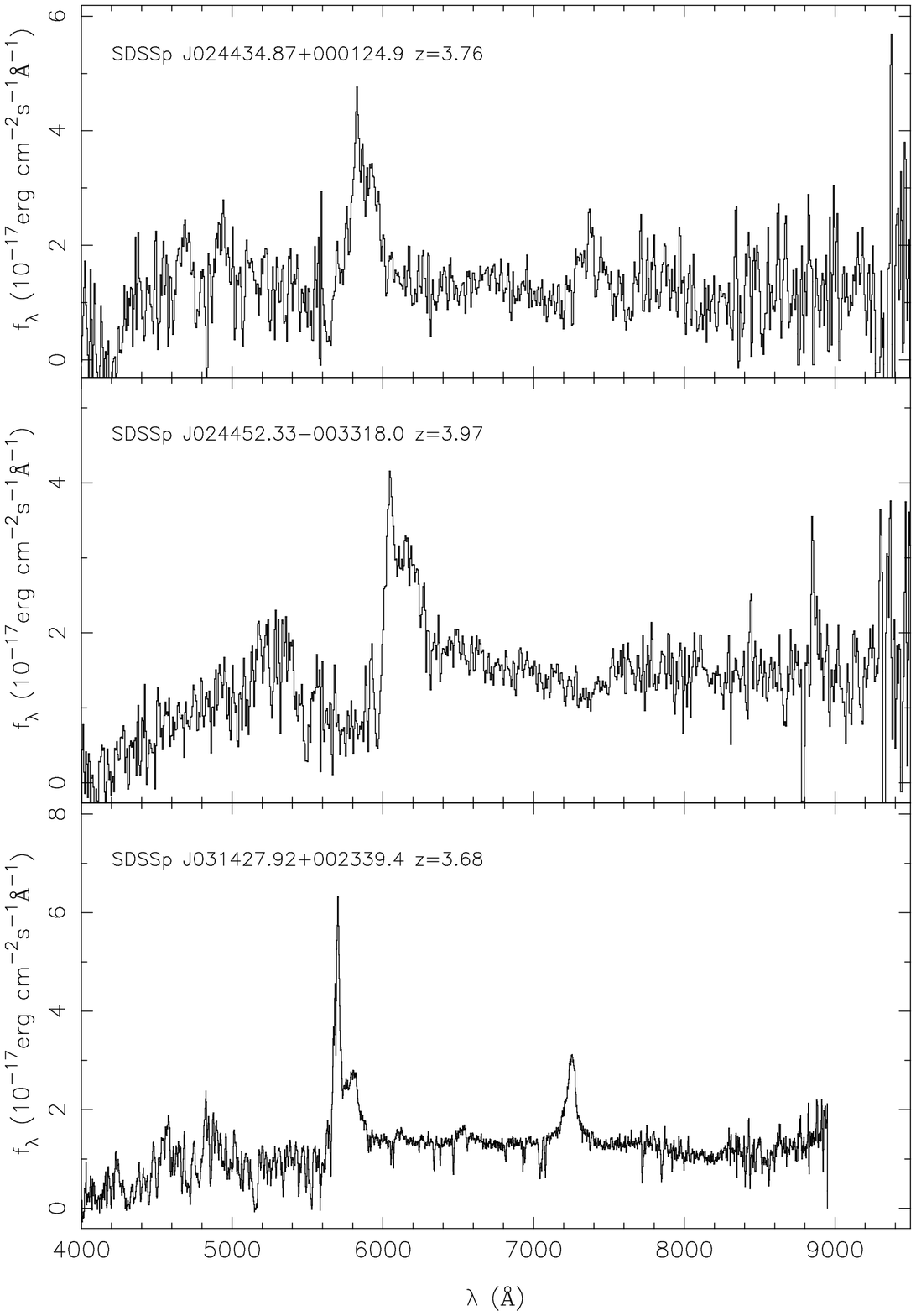}
\vspace{1cm}
Figure B2. Continued

\end{figure}
\newpage

\begin{figure}
\vspace{-3cm}

\plotone{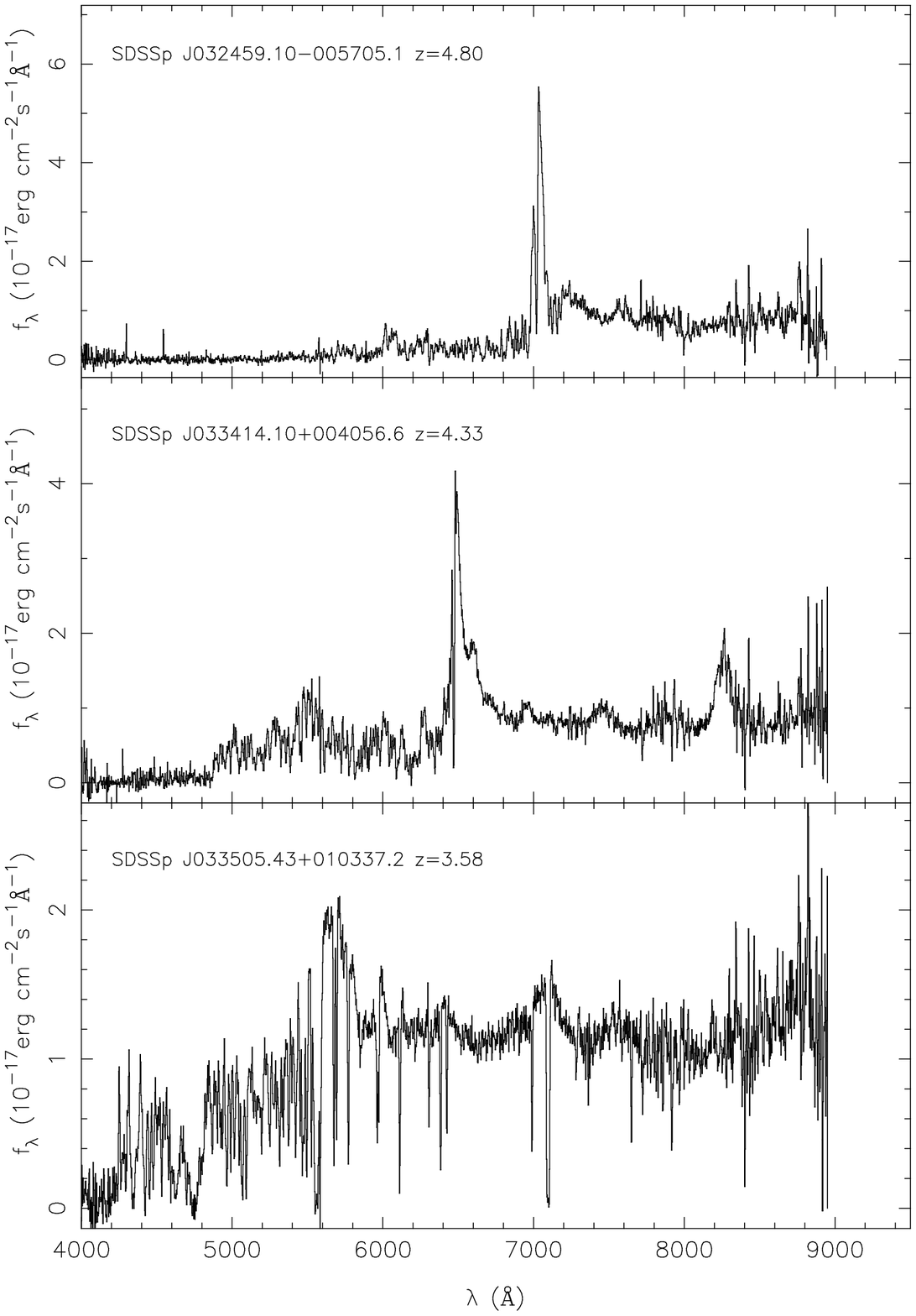}
\vspace{1cm}
Figure B2. Continued

\end{figure}

\newpage
\begin{figure}
\vspace{-3cm}

\plotone{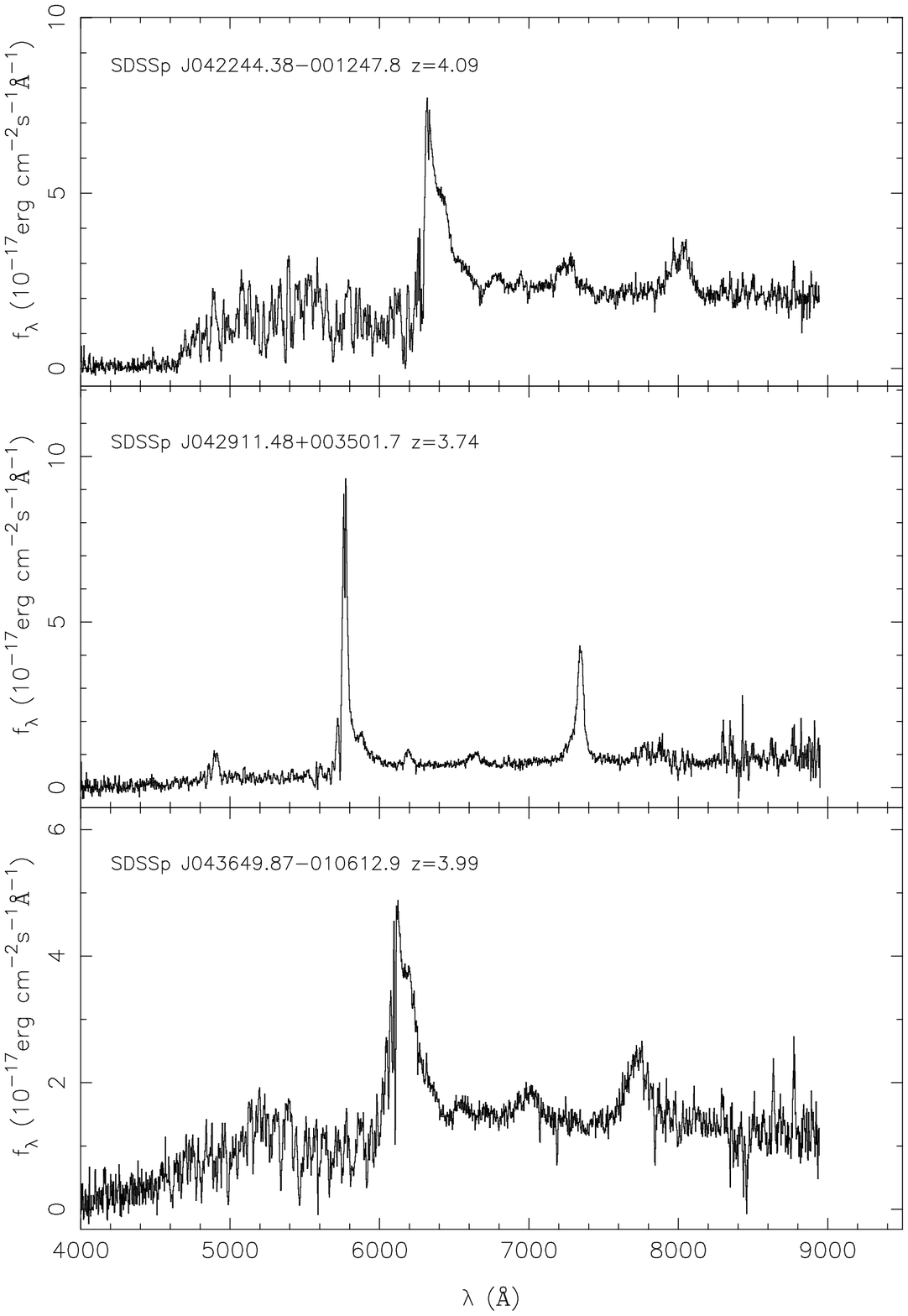}
\vspace{1cm}
Figure B2. Continued

\end{figure}

\newpage

\end{document}